\documentclass[12pt,preprint]{aastex}   

\newcommand{\Lsol}{\mbox{$L_\odot$}}
\newcommand{\Msol}{\mbox{$M_\odot$}}
\newcommand{\Vlsr}{\mbox{$V_{\rm LSR}$}}
\newcommand{\kms}{\mbox{km s$^{-1}$}}
\newcommand{\beam}{\mbox{beam$^{-1}$}}
\newcommand{\persquarecm}{\mbox{cm$^{-2}$}}
\newcommand{\percubiccm}{\mbox{cm$^{-3}$}}
\newcommand{\peryr}{\mbox{yr$^{-1}$}}

\newcommand{\HH}{\mbox{H$_2$}}
\newcommand{\minus}{\mbox{$-$}}
\newcommand{\xone}{\mbox{$x_{1}$}}
\newcommand{\xtwo}{\mbox{$x_{2}$}}

\newcommand{\twelveCO}{\mbox{$^{12}$CO}}
\newcommand{\thirteenCO}{\mbox{$^{13}$CO}}
\newcommand{\CeighteenO}{\mbox{C$^{18}$O}}
\renewcommand{\tilde}{\mbox{$\sim$}}
\newcommand{\hr}{\mbox{$^{\rm h}$}}
\newcommand{\mn}{\mbox{$^{\rm m}$}}
\newcommand{\cubiccm}{\mbox{cm$^{-3}$}}
\newcommand{\twotoone}{\mbox{(2--1)}}

\shorttitle{Molecular Superbubbles in NGC 253 }
\shortauthors{Sakamoto et al. }

\slugcomment{Accepted for publication in {\it The Astrophysical Journal}}

\begin{document}
\title{Molecular Superbubbles in the Starburst Galaxy NGC 253 }
\author{Kazushi Sakamoto\altaffilmark{1,2}, 
Paul T. P. Ho\altaffilmark{3}, 
Daisuke Iono\altaffilmark{1}, 
Eric R. Keto\altaffilmark{3}, 
Rui-Qing Mao\altaffilmark{4}, \\
Satoki Matsushita\altaffilmark{5}, 
Alison B. Peck\altaffilmark{1}, 
Martina C. Wiedner\altaffilmark{6}, 
David J. Wilner\altaffilmark{3}, \\
and 
Jun-Hui Zhao\altaffilmark{3}}
\altaffiltext{1}{Harvard-Smithsonian Center for Astrophysics,
Submillimeter Array, 645, N. A'ohoku Place, Hilo, HI 96720}
\altaffiltext{2}{National Astronomical Observatory of Japan,
Mitaka, Tokyo 181-8588, Japan}
\altaffiltext{3}{Harvard-Smithsonian Center for Astrophysics,
60 Garden Street, Cambridge, MA 02138}
\altaffiltext{4}{Purple Mountain Observatory, Chinese Academy of Sciences, Nanjing, 210 008, China}
\altaffiltext{5}{Academia Sinica, Institute of Astronomy and Astrophysics, 
P.O. Box 23-141, Taipei 106, Taiwan}
\altaffiltext{6}{I. Physikalisches Institut, Universit\"at zu K\"oln, Z\"ulpicher Str. 77,
50937 K\"oln, Germany}

\begin{abstract}
The central $2\times1$ kpc of the starburst galaxy NGC 253 has been imaged using the Submillimeter Array
at a 60 pc resolution in the J=2--1 transitions of \twelveCO, \thirteenCO, and \CeighteenO\ 
as well as in the 1.3 mm continuum.
Molecular gas and dust  are mainly in the circumnuclear disk of \tilde500 pc radius, 
with warm (\tilde40 K) and high area-filling factor gas in its central part.
Two gas shells or cavities have been discovered in the circumnuclear disk.
They have \tilde100 pc diameters and have large velocity widths of 80--100 \kms, 
suggestive of expansion at  \tilde50 \kms.
Modeled as an expanding bubble, each shell has an age of  \tilde0.5 Myr 
and needed kinetic energy of \tilde$1 \times 10^{46}$ J 
as well as mean mechanical luminosity of \tilde$1 \times 10^{33}$ W  for its formation.
The large energy allows each to be called a superbubble.
A \tilde$10^6$ \Msol\ super star cluster can provide the luminosity, 
and could be a building block of the nuclear starburst in NGC 253.
Alternatively, a hypernova can also be the main source of energy for each superbubble, 
not only because it can provide the mechanical energy and luminosity 
but also because the estimated rate of superbubble formation and that of hypernova explosions are comparable.
Our observations indicate that the circumnuclear molecular disk harboring the starburst is 
highly disturbed on 100 pc or smaller scales, 
presumably by individual young clusters and stellar explosions, 
in addition to globally disturbed in the form of the well-known superwind. 
\end{abstract}


\section{Introduction}
The center of the nearby spiral galaxy NGC 253 is one of the best places 
to study starburst activity in detail
because of its proximity \citep[$D=3.5$ Mpc;][]{Rekola05}.
A burst of star formation has been taking place in the central \tilde500 pc for a few $10^7$ years \citep{Rieke80,Engelbracht98},
producing the far-infrared luminosity of \tilde$3\times 10^{10}$ \Lsol\ in the shroud of dust \citep{Telesco80}.
At least part of the star formation has occurred in the form of compact star clusters 
with a few parsec radii and \tilde$10^{5}$ -- $10^{6}$ \Msol\ masses \citep{Watson96,Keto99}.
The young luminous clusters called super star clusters have been suspected to be  globular clusters in formation.
The starburst has resulted in the high rate of supernova explosions, \tilde$0.1$ \peryr, in the nuclear region \citep{Rieke88,vanBuren94,Ulvestad97}.
The rapid injection of energy into the interstellar medium has caused a kiloparsec-scale outflow
of hot gas, called a superwind, in the direction perpendicular to the galactic disk
\citep{Ulrich78,Fabbiano84,Tomisaka88,Heckman90}.

The fuel for the starburst is the abundant molecular gas in the central kiloparsec,
apparently accumulated with the help of a 7 kpc-long stellar bar \citep{Scoville85,Mauersberger96}.
Higher-resolution imaging with millimeter arrays showed 
that molecular gas in the starburst region is distributed 
in a bar-like \tilde$0.7\times0.2$ kpc structure containing several dense clumps, 
presumably formed in response to the dynamical perturbation from the stellar bar
\citep{Canzian88,Carlstrom90,Paglione95, Peng96, Garcia-Burillo00,Paglione04}. 
The properties of the molecular gas in the region are of great importance
to further understand the mechanisms and evolution of the starburst,
because the gas properties affect the formation of stars and are
affected by the newly formed stars.

To further study this nearby archetypal starburst,  
we observed the molecular gas and dust in the nucleus of NGC 253 at 1.3 mm 
using the Submillimeter Array (SMA)\footnote{
The Submillimeter Array is a joint
project between the Smithsonian Astrophysical Observatory and the
Academia Sinica Institute of Astronomy and Astrophysics, and is
funded by the Smithsonian Institution and the Academia Sinica.
}
on the summit of Mauna Kea, Hawaii.
This new telescope at 4080 m operates above most of the water in the
atmosphere, and allowed us to image with high sensitivity and resolution,
revealing new features in the starburst region.
In this paper, we report the SMA observations of the nuclear region with special emphasis on our discovery 
of two molecular shells that can be best modeled as molecular superbubbles.
The superbubbles, which are the first ones found in NGC 253, 
are suggested to be the end products of active star formation.
We introduce our SMA observations in \S2, 
describe the overall properties of gas and dust in the central kiloparsec in \S3,
derive the properties of the expanding molecular superbubbles in \S4,
discuss their formation mechanism and evolution in \S5,
and summarize our findings in \S6.

\begin{figure}
\epsscale{0.6}
\plotone{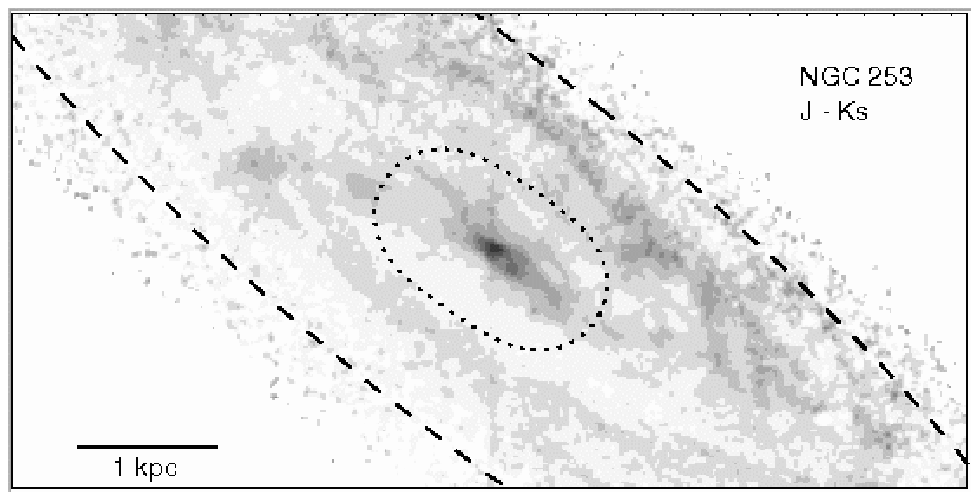} \\
\epsscale{0.6}
\plotone{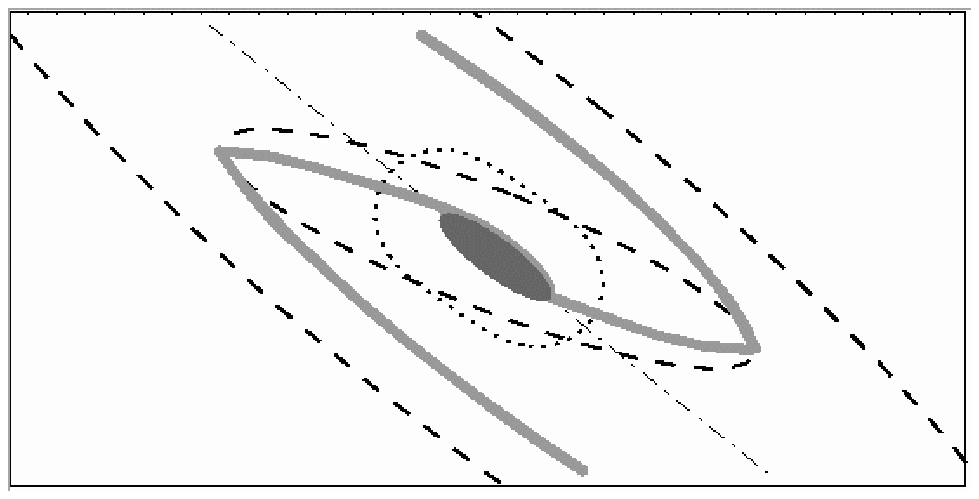} 
\caption{(top) The $J-K_{\rm s}$ color index image of NGC 253 made from the 2MASS data 
of \citet{Jarrett03}.
The gray scale is from 0.9 mag to 2.3 mag with darker regions being redder, or showing higher
extinction by interstellar dust. 
The dashed ellipse, only a part of which is in the frame, is the galaxy's outline at $K_{\rm s}=20$ mag.
The dotted oval in the center shows the field of view of our mosaic CO observations.
North is up and east is to the left. 
The northwest side, i.e., the top-right side, is the near side of the galaxy.
(bottom) An illustration showing the main features in the galaxy.
The new dashed ellipse is the stellar bar that is 7 kpc long on the plane of the galaxy. 
The gray thick curves delineate dust lanes in the leading edge of the stellar bar 
as well as the spiral arms starting at the ends of the bar.
The gray ellipse in the center represents the circumnuclear disk of molecular gas where a burst of
star formation has been taking place.
The dash-dotted line is the line of nodes.
 \label{fig.illustration} }
\end{figure}

We adopt the above-mentioned distance of 3.5 Mpc ($1''=17$ pc) for the galaxy,
and scale the distance-dependent values in the literature accordingly.
Also adopted are the following parameters;
78\degr\ for the inclination of the galaxy  \citep{Pence80},
51\degr\  for the position angle (P.A.) of the approaching line of nodes \citep{Burbidge62,Jarrett03},
and
\tilde68\degr\ for the P.A. of the large scale stellar bar in the galaxy \citep{Scoville85}.
The northwest side of the galaxy is the near side to us, assuming that the
spiral arms in the galaxy are trailing.
Figure \ref{fig.illustration} helps readers to grasp the configuration and to locate the
region we discuss in this paper. 
The center of the galaxy is assumed to be 
at $\alpha$=00\hr47\mn33\fs17,  $\delta$=\minus25\arcdeg17\arcmin17\farcs1 (J2000),
as has been usually adopted in the literature.
The position is determined from the brightest compact radio source in the galaxy, 
possibly an AGN \citep{Turner85,Ulvestad97}.
The radio source was suggested to coincide with the galactic center at  sub-arcsecond accuracy 
on the basis of the morphology in the infrared  and the kinematics of dense molecular gas  
\citep{Kalas94,Garcia-Burillo00}.

\section{Observations and Data Reduction}
The central $2\arcmin \times 1\arcmin$ (\tilde$2 \times 1$ kpc on the sky) of NGC 253 was observed 
with the Submillimeter Array on the 19th and 21st November 2003, 
during the commissioning of the new submillimeter interferometer 
\citep[see][for description of the telescope]{Ho04}.
The eight 6-meter antennas of the array were in a compact configuration 
that provided projected baselines ranging from 7 m to 75 m, which
correspond to spatial frequencies from 4\arcsec\ to 40\arcsec.
We observed three positions 26\arcsec\ apart from each other; 
the interval is the half width at half maximum (HWHM) of the primary beam. 
The central position was at  the galactic center and the two outer positions were at position angles of 
55\degr\ and 235\degr, almost on the major axis of the galaxy.
The quasar J0132\minus169 was observed every 30 minutes for gain calibration, Jupiter and Saturn were observed
for passband calibration, and Uranus was used for absolute flux calibration.
Interferometric pointing checks were made toward the quasar J2258\minus279 
a few times before and during the observing run, and toward Uranus before flux calibration.
Pointing offsets, which typically changed by a few arcsec, were updated about an hour after each
pointing measurement because J2258\minus279 precedes the galaxy by two hours.
The SIS receivers were tuned to simultaneously observe three CO lines in the 1 mm band:
\twelveCO(2--1) in the upper sideband, and \thirteenCO(2--1) and \CeighteenO(2--1) in the
lower sideband. 
The median double-side-band system temperature was 200 K toward the galaxy.
The digital correlators having a 2 GHz bandwidth were configured for 0.8125 MHz spectral
resolution.
We lost one antenna in the first observing run 
and lost \twelveCO\ data from two antennas in the second observing run  
because of an antenna and correlator fault, respectively.
The total integration time on NGC 253 was 7.5 hours.

Data reduction was made using MIR, which is an IDL version of MMA \citep{Scoville93a}, 
MIRIAD \citep{Sault95}, and the NRAO AIPS package \citep{Bridle94}. 
The data were calibrated for the passband, 
corrected for atmospheric absorption using the system temperatures,
and then calibrated for the complex gain using the calibrator data.
The flux density of the calibrator J0132\minus169 was estimated to be 
1.50 Jy in LSB (220 GHz) and 1.44 Jy in USB (230 GHz) from the comparison with Uranus,
for which we adopted the brightness temperature given by \citet{Griffin93}. 
Continuum data were obtained from channels that do not have line emission; 
the resulting data have an effective bandwidth of 0.7 GHz in each sideband.
The continuum was then subtracted from the visibility data of the line emission.
Channel maps for the line emission were made with  5 \kms\ resolution for \twelveCO\ 
and 10 \kms\ resolution for \thirteenCO\ and \CeighteenO.
The rms noise is 124 mJy \beam\ (0.26 K), 44 mJy \beam\  (0.10 K),
and 39 mJy \beam\ (0.08 K) in the \twelveCO, \thirteenCO, and \CeighteenO\  channel maps, respectively,
and is 4 mJy \beam\ in the continuum map made by combining the two sidebands.
The synthesized beam is almost the same among the lines and continuum, ranging
from 4\farcs3 $\times$ 2\farcs5 to 4\farcs8 $\times$ 2\farcs7.
The spatial resolution at the distance of the galaxy is about 60 pc.
We mosaiced the  images using the so-called joint approach, in which
maps from individual pointing positions were combined first
in order to use information from all pointing positions in the subsequent deconvolution.
Comparison with single-dish observations indicates that our interferometric
observations recovered about 90 \% of the total flux in each line as well as in continuum.
Specifically, we detected, 91\%, 116 \%, and 89 \% of 
the single-dish integrated intensities of the \twelveCO, \thirteenCO, and \CeighteenO\ emission, 
respectively, obtained in the central 23\arcsec\ of the galactic center
by \citet{Harrison99}.
We also detected 84 \% of the single-dish flux density of 1.3 mm continuum
that was obtained by \citet{Krugel90} and is corrected by 30 \% for 
line contamination \citep{Mauersberger96}.
Velocities in this paper are with respect to the Local Standard of Rest (LSR) 
and are defined in the radio convention. 
Unless otherwise noted, maps are shown without correcting for the attenuation by the mosaiced primary beam, 
though all the flux and intensity measurements were made after the correction.

\section{Overall Properties of Gas and Dust in the Central kiloparsec}
The overall distribution of molecular gas and dust in the starburst nucleus of NGC 253 is
shown in the maps of the CO lines and 1.3 mm continuum in Fig. \ref{fig.total}.
The main component is a 50\arcsec\ (0.9 kpc) long linear feature 
elongated in the position angle of about 58\degr.
The feature was first recognized as a molecular gas bar by \citet{Canzian88}
and was subsequently modeled as a highly inclined disk
near the center of the 7 kpc stellar bar \citep[e.g.,][]{Peng96,Garcia-Burillo00, Paglione04}.
We simply call the gas structure a (circum)nuclear gas disk. 
The nuclear gas disk extends from about $-20\arcsec$ to $+30\arcsec$ in the R.A. offset
in the \twelveCO\ map.
Weaker CO emission extends further in the direction of the stellar bar.

\subsection{Continuum Emission}
The total flux density of the 1.3 mm continuum is $1.0\pm 0.1$ Jy in our data, 
whereas that in previous 3.0--3.5 mm observations is 0.25--0.35 Jy 
\citep{Carlstrom90,Peng96, Garcia-Burillo00}.
The 3 mm continuum contains roughly equal amounts of synchrotron, free-free, and
thermal dust emission \citep{Garcia-Burillo00}. 
The increased flux density at 1 mm must be from the dust emission 
because it has a positive spectral slope
as a function of frequency unlike synchrotron and optically thin free-free emission.
Thus most ($\gtrsim 80$ \%) of the 1.3 mm continuum must be thermal emission from dust.

The continuum emission peaks at 
$\alpha$=00\hr47\mn33\fs18,  $\delta$=\minus25\arcdeg17\arcmin17\farcs4 (J2000)
and is elongated along the nuclear gas disk.
The position of the peak agrees, within \tilde1/10 of our synthesized beam, with the adopted galactic center,
where the brightest centimeter source in the galaxy resides.
The radio source called TH2 has been suggested to be an AGN 
\citep{Turner85, Mohan02}.
The millimeter continuum	shows a possible secondary peak at
$\alpha$=00\hr47\mn32\fs96,  $\delta$=\minus25\arcdeg17\arcmin19\farcs9 (J2000),
which is about 4\arcsec\ southwest of the primary peak and is within 0\farcs4 from another
radio source (called TH8) listed by \citet{Turner85}.
In the mid-infrared, the brightest peak in the nuclear region is a super star cluster 3\arcsec\
southwest of the galactic center \citep{Keto99,Boeker98}. 
The cluster is located between the two millimeter peaks.

\subsection{Molecular Line Emission}
It is known that the bulk of \twelveCO(2--1) emission is optically thick 
while the \thirteenCO(2--1) and \CeighteenO(2--1) lines are optically thin in the central
region of NGC 253, 
with the mean optical depths in \twelveCO(2--1) and \thirteenCO(2-1) being 
\tilde3 and \tilde0.08 respectively  \citep{Henkel93,Harrison99}.
These opacities are based on single-dish observations at 13\arcsec--23\arcsec\ resolutions.
The \thirteenCO\ line is still optically thin at the higher spatial resolution of our data.
The maximum value of the \thirteenCO/\twelveCO\  brightness temperature ratio is found to be 
0.4 in our data, after correcting for the slightly different spatial resolution
and primary beam attenuation.
This ratio corresponds to the \thirteenCO(2--1) optical depth of 0.5,
i.e., $\max(\tau[\thirteenCO\twotoone]) \sim 0.5$,
under the assumptions 
of LTE and the same beam filling factor for the two lines.
While \thirteenCO(2--1) is still optically thin even at the most opaque regions at our current resolution,
\twelveCO(2--1) has a very high opacity 
in the same regions and suffers from self absorption;
$\max(\tau[\twelveCO\twotoone]) \sim 20$, assuming  $^{12}$C/$^{13}$C \tilde\ 40 
estimated for NGC 253 by \citet{Henkel93}.
The same analysis for \CeighteenO\ results in $\max(\tau[\CeighteenO\twotoone]) \sim 0.15$.

Molecular gas in the nuclear gas disk is warm.
For example, the peak brightness temperature observed in our \twelveCO(2--1) data cube is 36 K;
the physical temperature that would produce the same intensity with the
Planck formula is 41 K.
The peak intensity is significantly higher than that of the same line in the central kiloparsec of our Galaxy.
The peak brightness temperature there is 11 K when the  \twelveCO(2--1) observations 
by \citet{Sawada01} are convolved to our resolution.
The presence of warm gas inferred from the peak brightness temperature is consistent with
previous observations of NGC 253.
The excitation analyses of multiple transitions of various molecular lines, such as CO, HCN, and NH$_{3}$,
have indicated that the bulk of molecular gas  is warm, \tilde50 --  120 K, in the central \tilde300 pc of the galaxy
\citep{Gusten93, Jackson95, Mauersberger03, Bradford03}.

\subsection{Structure of the Circumnuclear Gas Disk}
We infer three properties of the circumnuclear gas disk from our
data, namely the flaring of low density gas from the galactic plane, 
a high area-filling factor of molecular gas toward the center of the disk,
and the distinct  characteristics of the central 100 pc.

\begin{figure}
\epsscale{1.0}
\plottwo{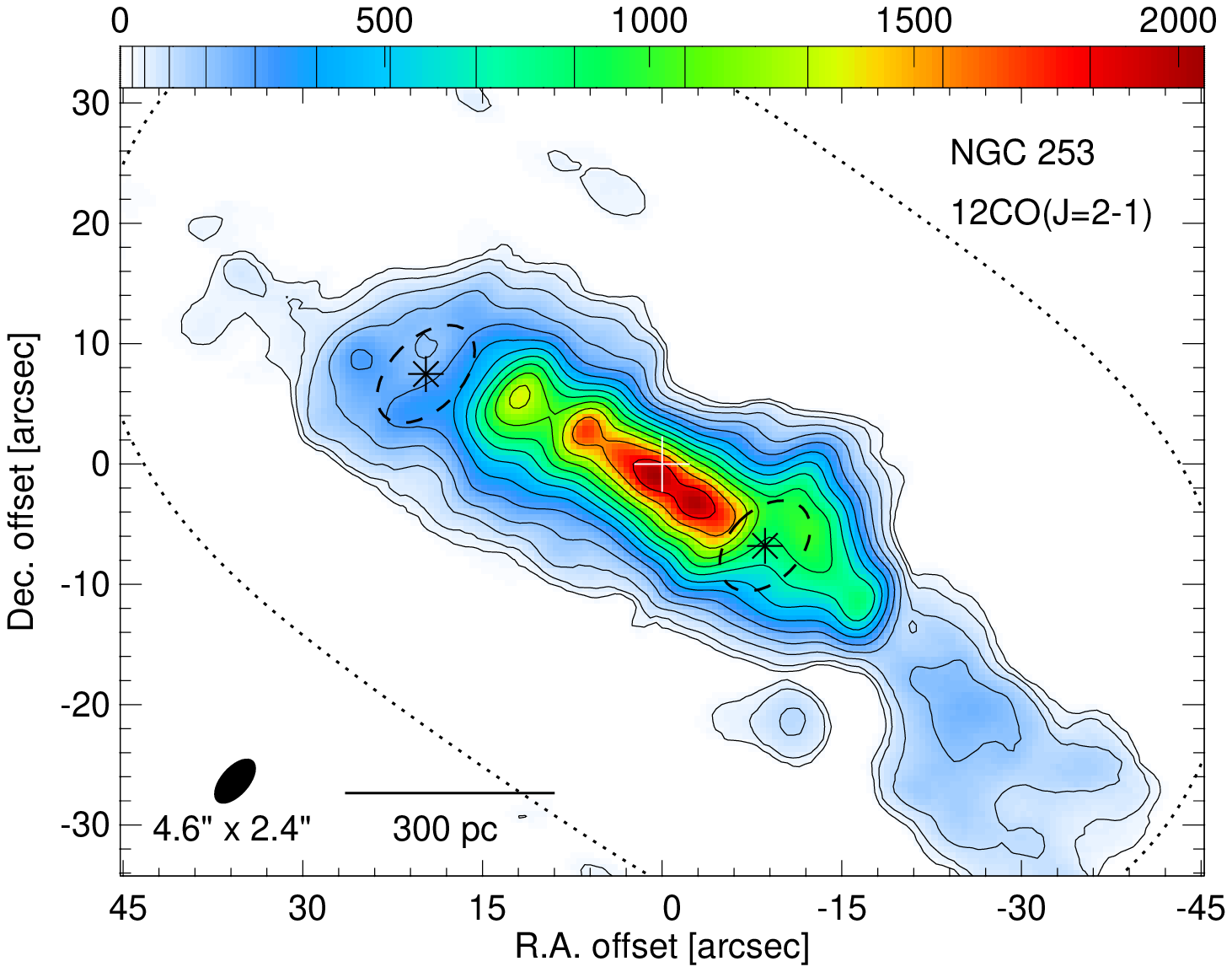}{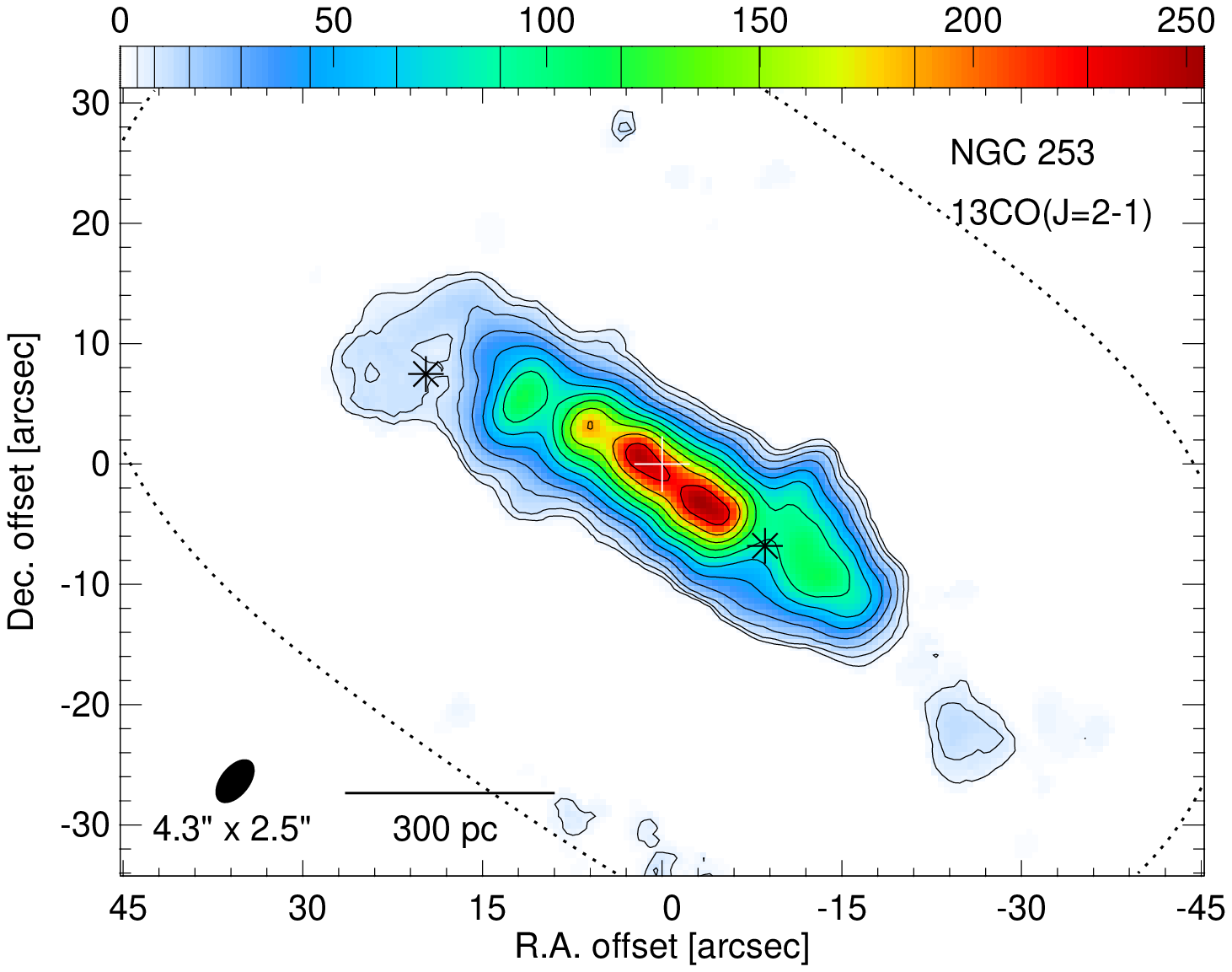} \\
\epsscale{1.0}
\plottwo{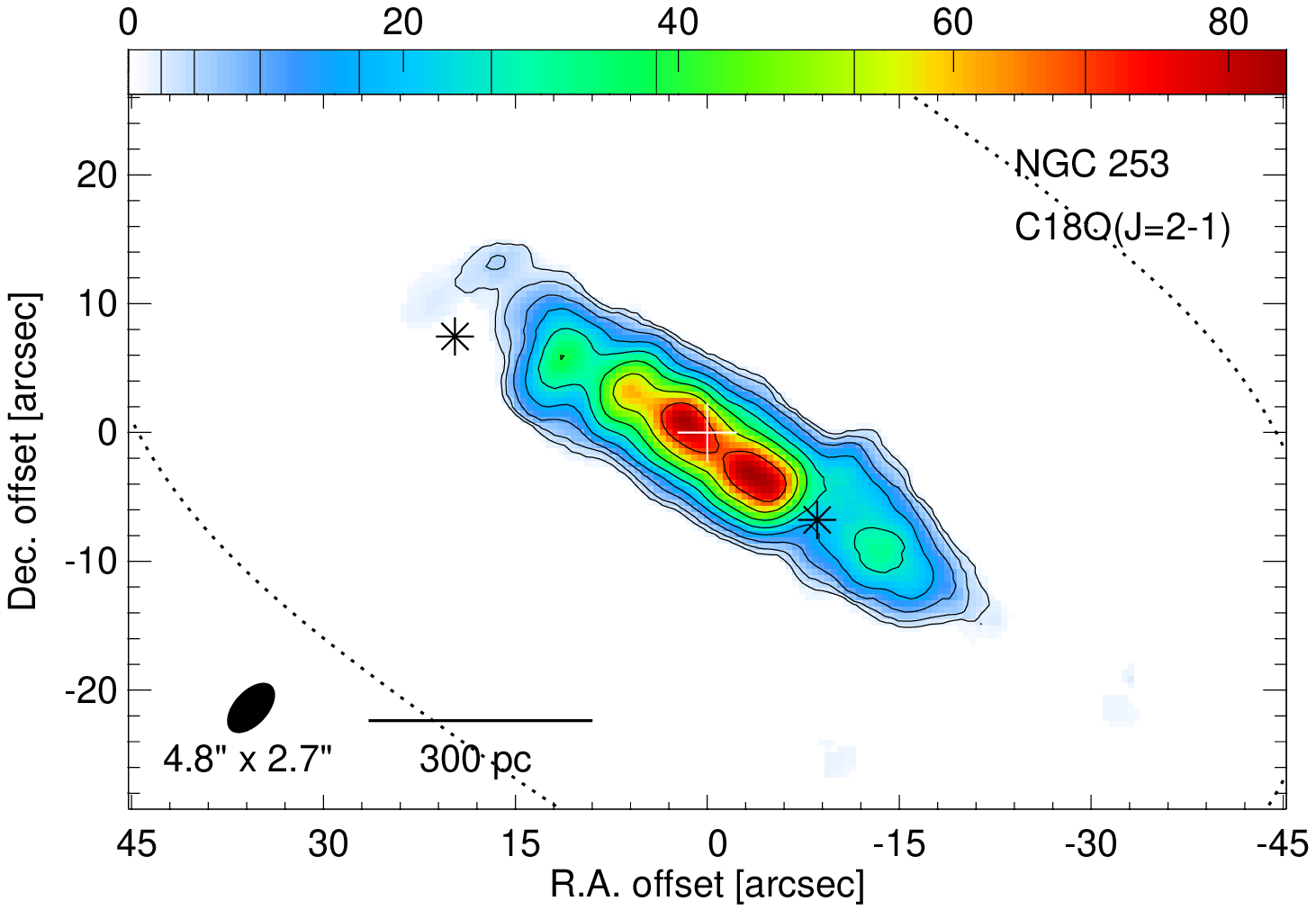}{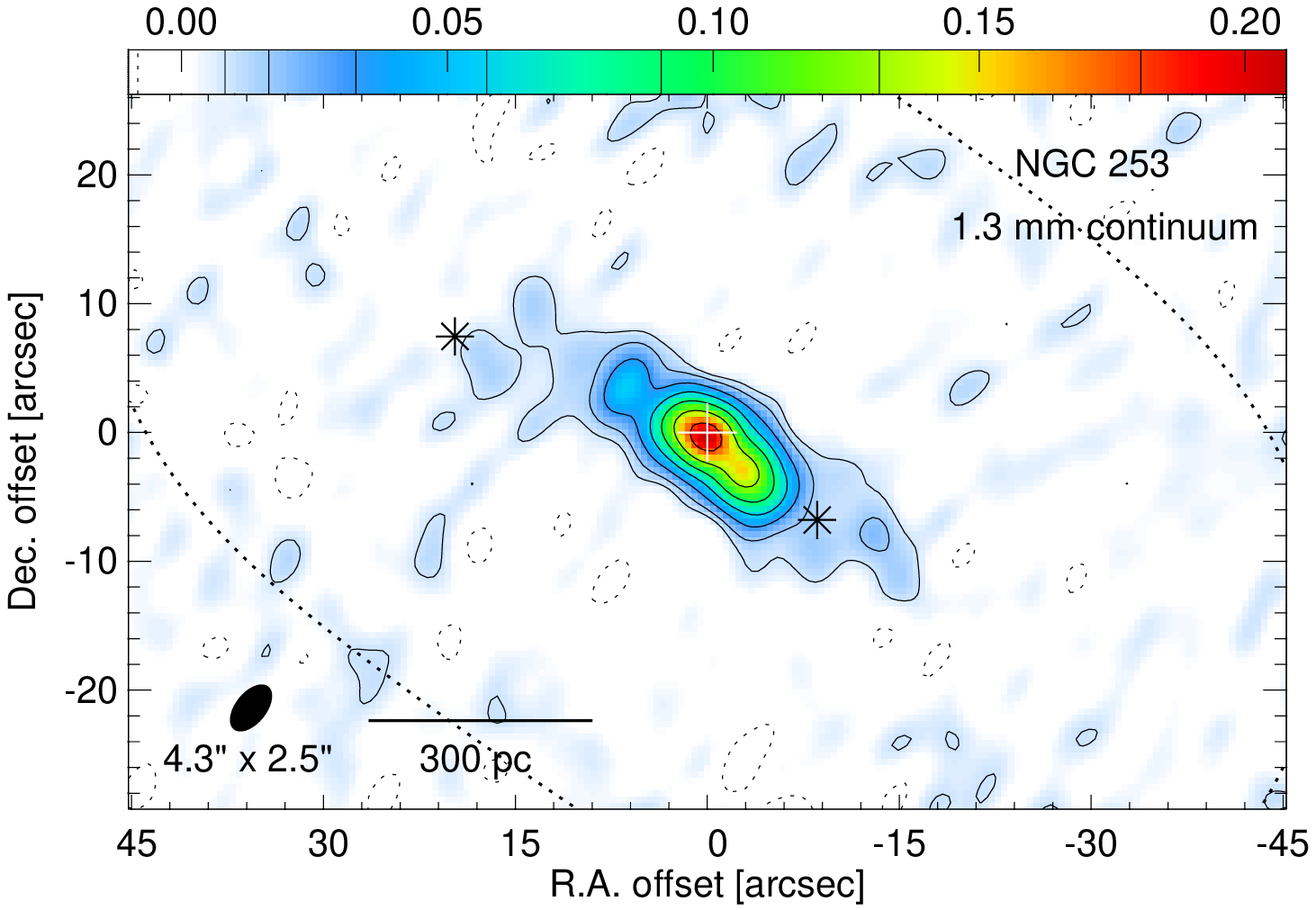}
\caption{The starburst nucleus of NGC 253 in \twelveCO(2--1),  \thirteenCO(2--1), \CeighteenO(2--1), 
and 1.3 mm continuum.
Positional offsets are measured from 
$\alpha$=00\hr47\mn33\fs17,  $\delta$=\minus25\arcdeg17\arcmin17\farcs1 (J2000),
which is the assumed position of the galactic center and is shown as a cross.
The brightest centimeter-wave source in the galaxy, called TH2, is at this position.
The unit of intensity is Jy \beam\ \kms\ for the CO integrated intensity maps and Jy \beam\ for the continuum map,
which is made by averaging the USB and LSB maps.
The FWHM of the synthesized beam  is shown at the bottom left corner of each map, 
and the mosaiced primary beam is indicated by a dotted line at 40 \% of its peak.
The positions of the molecular superbubbles are shown with asterisks, 
and their approximate sizes are shown with ellipses in the \twelveCO\ image.
The superbubble called SB1 is to the southwest and the one called SB2 is to the
northeast of the nucleus.
 \label{fig.total} }
\end{figure}

The flaring of low density gas is inferred from the
observation that the \twelveCO\ distribution looks 
wider compared to its length than the \thirteenCO\ and \CeighteenO\  distributions (see Fig. \ref{fig.total}).
The observation was confirmed by fitting an elliptical Gaussian to our data.
The deconvolved minor-to-major axial ratio is 0.24 in \twelveCO, while
it is 0.19 and 0.18 for \thirteenCO\ and \CeighteenO, respectively.
The fitting used maps made by simply summing channel maps 
rather than the moment maps in Fig.  \ref{fig.total}, in order to avoid the bias 
caused by the intensity cutoff in the moment analysis.
The attenuation of the primary beam was corrected before the fitting.
The fractional uncertainty of the axial ratio is largest for \twelveCO\ and is 2 \% for the line,
apparently because its distribution is dumbbell-shaped to some extent.
For comparison, the axial ratio expected  from the inclination of the galaxy is 0.20 
for a thin uniform disk.
The broadening in \twelveCO\ may be due to the presence of tenuous molecular gas 
above and below the plane of the galaxy, possibly as part of the superwind. 
If the flared gas has a density slightly below the critical density of the
CO(2--1) line (i.e., if $n_{\rm H_2}\approx10^{3}$ -- $10^{4}$ \cubiccm), 
it could effectively emit  only in \twelveCO\ with the help of
photon trapping.
Entrainment of ambient ISM by the superwind has been suggested 
from extended PAH emission along the minor axis \citep{Tacconi-Garman05}.

\begin{figure}
\epsscale{0.8}
\plotone{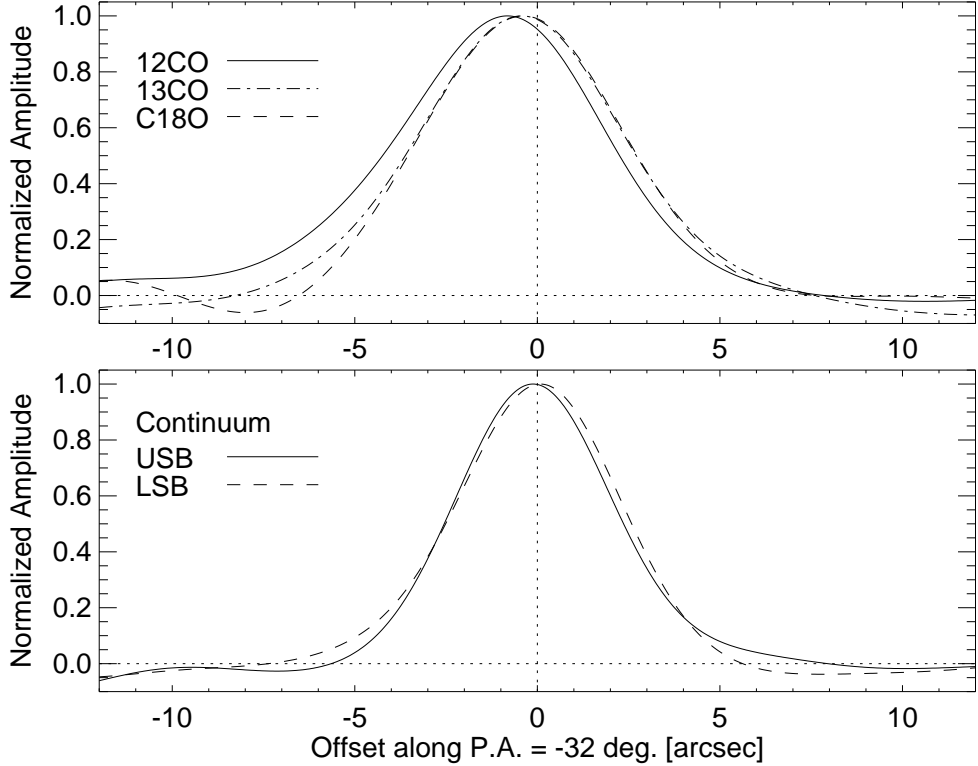} \\
\caption{The intensity profiles of \twelveCO(2--1),  \thirteenCO(2--1), \CeighteenO(2--1), 
and 1.3 mm continuum across the continuum peak along the position angle of $-32\degr$,
which is approximately the minor axis of the nuclear gas disk.
The profiles are measured from the maps that are corrected for the primary beam attenuation and
convolved to the same spatial resolution (the resolution of the \CeighteenO\ map).
The CO line maps are made by summing up channel maps rather than by moment analysis. 
The amplitude of each emission is normalized so that the peak intensity is unity.
Note that each emission is only partially resolved in this direction.
The peak signal-to-noise ratio is 330, 75, 29, 29 and 39 for \twelveCO,  \thirteenCO, \CeighteenO,
and USB and LSB continuum, respectively.
The \twelveCO(2--1) profile is 0\farcs5 offset from that of \thirteenCO(2--1) and \CeighteenO(2--1).
 \label{fig.slice} }
\end{figure}

The high area-filling factor of molecular gas in the central part of the nuclear disk is implied 
by the small offset of the \twelveCO\ ridge toward the southeast 
from the \thirteenCO\ and \CeighteenO\ ridges as well as from the continuum peak.
Figure \ref{fig.slice} shows the intensity profiles of those lines and continuum 
through the continuum peak at the position angle of 238\degr.
It confirms the offset of 0\farcs5 between \twelveCO\ and
its optically thin isotopomers.
The offset is unlikely to be an artifact of a relative phase-calibration error  between the two
sidebands, judging from the
agreement of USB and LSB continuum profiles in the lower panel of Fig. \ref{fig.slice}.

A plausible model for the offset is that 
the molecular disk with a finite thickness has a high-area filling factor in its central part.
In this model, we see the near-side surface of the disk in \twelveCO\ 
while we see through the disk in optically-thin emission.
For a disk of half-height $h$ and inclination $i$, 
the offset between \twelveCO\ and optically thin tracers will be $h \sin i$. 
The observed offset thus corresponds to the 20 pc thickness (=$2h$) of the nuclear disk 
measured at its \twelveCO\ photosphere. 
The disk can be made of gas clumps, 
but they must be so crowded in sky position and in line-of-sight velocity 
that clump overlap is significant near the center of the disk. 

The high area-filling factor (and hence little beam dilution) in the center of the disk  suggests
that the peak brightness temperature of \tilde40 K is a good estimate for 
the mean physical temperature of molecular gas at the photosphere. 
For example, it is unlikely that there are only  small clouds of 200 K filling 1/5 of the beam area. 
The model also suggests that a central hole in the circumnuclear disk, if any, is insignificant at our current resolution. 
If, for example, the nuclear disk were devoid of molecular gas in the central few 100 pc 
as is often observed in the center of barred galaxies \citep[e.g.,][]{Sakamoto04}, 
then that should show as a clear central depression in optically thin emission. 
Without such a feature, the nuclear disk must have molecular gas (with a high-area filling factor) in its center, 
though the gas disk could still have a ring-like region of high density around the galactic center 
as suggested from some observations \citep{Israel95,Garcia-Burillo00,Weaver02}.

The distinct characteristics of the central 100 pc of the nuclear gas disk is
indicated by the markedly different spatial distribution of the 1.3 mm continuum and that
of \thirteenCO\ and \CeighteenO\ integrated intensities.
The former is more strongly peaked toward the galactic center than the latter.
This may be because free-free emission, which is less than 20 \% of the 
1.3 mm continuum, is strongly peaked toward the galactic center.
Indeed the 1.3 cm radio map of \citet[see their Fig. 2]{Ulvestad97} would look like
the 1.3 mm map near the galactic center if convolved to the same resolution;
the brightest peak is at TH2 and the main extension is to the southwest.
If the free-free emission is the cause of the strong continuum peak at the galactic center, then
that means the  star formation is most intense there. 
Another class of possible reasons for the discrepancy 
between continuum and  \thirteenCO\ as well as \CeighteenO\ is the variation of
the properties of the neutral ISM near the galactic center.
The intensity of 1.3 mm dust emission should approximately be
the product of  dust mass opacity coefficient, the dust column density, and the temperature of the dust,
or
$  I_{\rm dust, 1.3\; mm} \propto \kappa_{\rm 1.3\; mm} \; \Sigma_{\rm dust} \; T_{\rm dust} $.
The intensity of an optically thin CO(2--1) line is approximately proportional to
the column density of molecular gas and the fractional abundance of the CO molecule with respect to \HH,
or
$I_{\rm CO(2-1)} \propto \Sigma_{\rm H_2} ({\rm[CO]/[H_2]})$,
if the density and temperature of the gas are 
in the range of 40--200 K and $10^{3}$ -- $10^{5.5}$ \cubiccm, respectively  \citep{Wild92}.
The bulk of molecular gas in the nuclear disk satisfies the condition \citep{Harrison99,Paglione04}.
Thus, if non-thermal emission is not strongly peaked toward the galactic center,
possible reasons for the higher continuum-to-\thirteenCO\ (or \CeighteenO) ratio 
within the central 100 pc of the nuclear gas disk include a higher temperature of the ISM,
higher dust-to-gas ratio, and the variation of the dust opacity coefficient 
or that of the CO isotopomer abundance.  
Each of these cases again suggests that the central 100 pc has distinct characteristics 
in the nuclear gas disk, likely due to the intense star formation there.

\section{Expanding Molecular Superbubbles}
\begin{figure}[t]
\epsscale{0.9}
\plotone{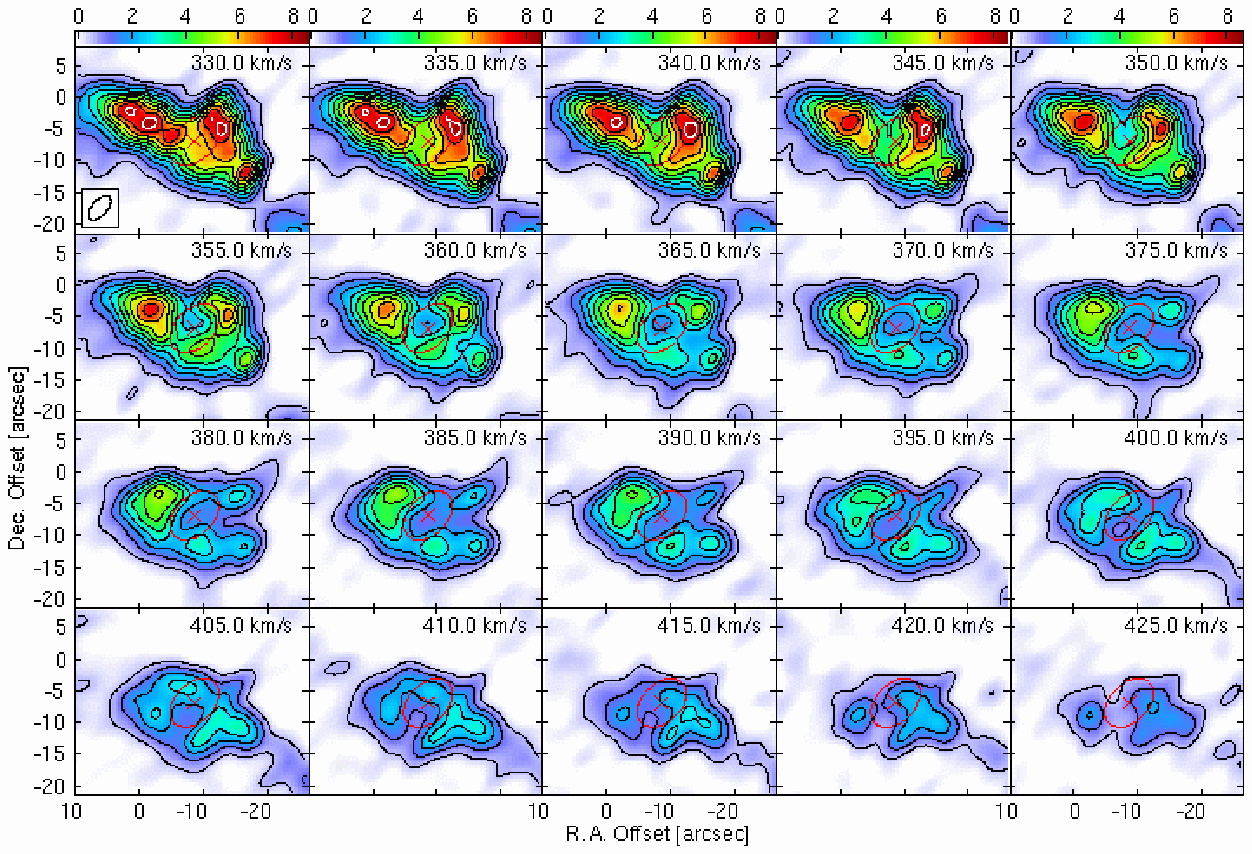}\\
\plotone{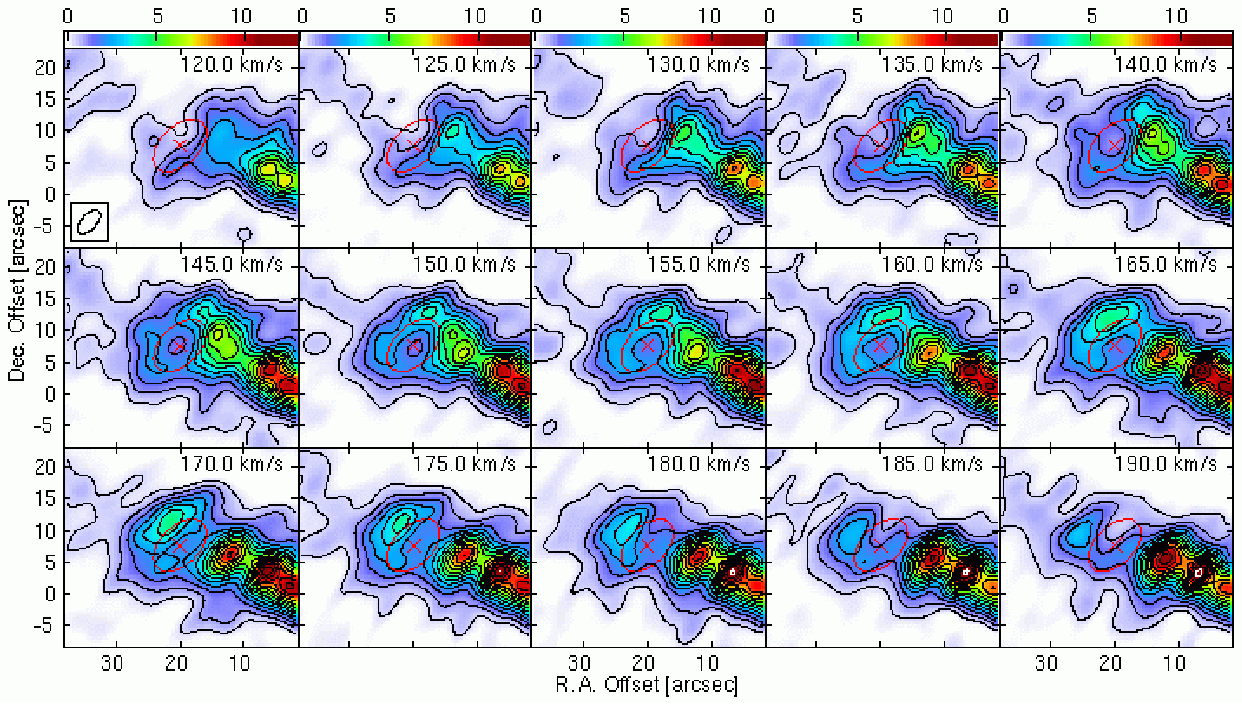}
\caption{\small \twelveCO(2--1) channel maps around the superbubble SB1 (top) and SB2 (bottom).
The $n$-th contour is at $n^{1.2} \times 0.5 $ Jy \beam\ (=4$\sigma$) for SB1
and $n^{1.3} \times 0.37$ Jy \beam\  (=3$\sigma$) for SB2.
The conversion to excess brightness temperature is 1 Jy \beam\ = 2.1 K.
The centroid position and approximate size of each bubble are marked with a red cross and a red
ellipse, respectively. 
Velocity is at the top-right corner of each panel, 
and the synthesized beam is shown at the bottom-left corner of the first panel.
 \label{fig.bubbles-channel} }
\end{figure}

	The \twelveCO\ channel maps in Fig. \ref{fig.bubbles-channel}
show two shell-like features in the nuclear gas disk.
A similar shell-like feature of comparable size 
has been observed in the starburst nucleus of M82, and has been interpreted
as an expanding molecular superbubble 
\citep{Neininger98, Wills99, Weiss99, Matsushita00, Weiss01, Matsushita05}.  
We hence call 
the one around $\Vlsr \sim 380$ \kms\  SB1 and another one
around  $\Vlsr \sim 150$ \kms\  SB2, under the working assumption that they are 
molecular superbubbles.
Both features are also seen in our \thirteenCO\ data, and SB1 is barely visible even in the
\CeighteenO\ data. 
SB1 is at a kink of the almost linear gas distribution (see Fig.   \ref{fig.total}), 
and may be responsible for it as we discuss below.
The kink was already seen in the first millimeter-interferometer image of the nuclear gas disk \citep{Canzian88},
though the shell feature there and another one on the other side of the nucleus
were not recognizable in previous observations because of the lack of sensitivity or
spatial resolution or both.
In the following, we derive the parameters of the shells from our data 
and summarize them in Table \ref{table.bubblesl}.

\subsection{Size, Shape, and Location}
The size of the shells, estimated from the channel maps and shown in the maps,
is about $160 \times 100$ pc.
The shells look elongated in the southeast-northwest direction, 
roughly along the minor axis of the galaxy.
The exact size and shape of the features are, however, difficult to determine at the current resolution.
Part of the elongation must be due to the elongation of the synthesized beam in the same direction. 
Also, it is almost certain that the shells have more complex shapes than an ellipse in the plane of the sky. 
For example, the shells may not be closed but have openings near its major axis 
as hinted in the channel maps, such as 350, 375, and 410 \kms\ maps of SB1.
These openings might indicate a conical or cylindrical shape elongated 
along the minor axis of the galaxy, as would be expected for galactic bubbles.  
It is also noted that the channels from around 370 \kms\ to 400 \kms\ show an emission tongue 
that appears to emanate from the area of SB1 by up to 10\arcsec\ (170 pc) in the minor axis direction.
Given the large inclination of NGC 253, the tongue emission is probably extended
in the direction perpendicular to the disk rather than within the plane of the galaxy.

\begin{figure}
\epsscale{0.6}
\plotone{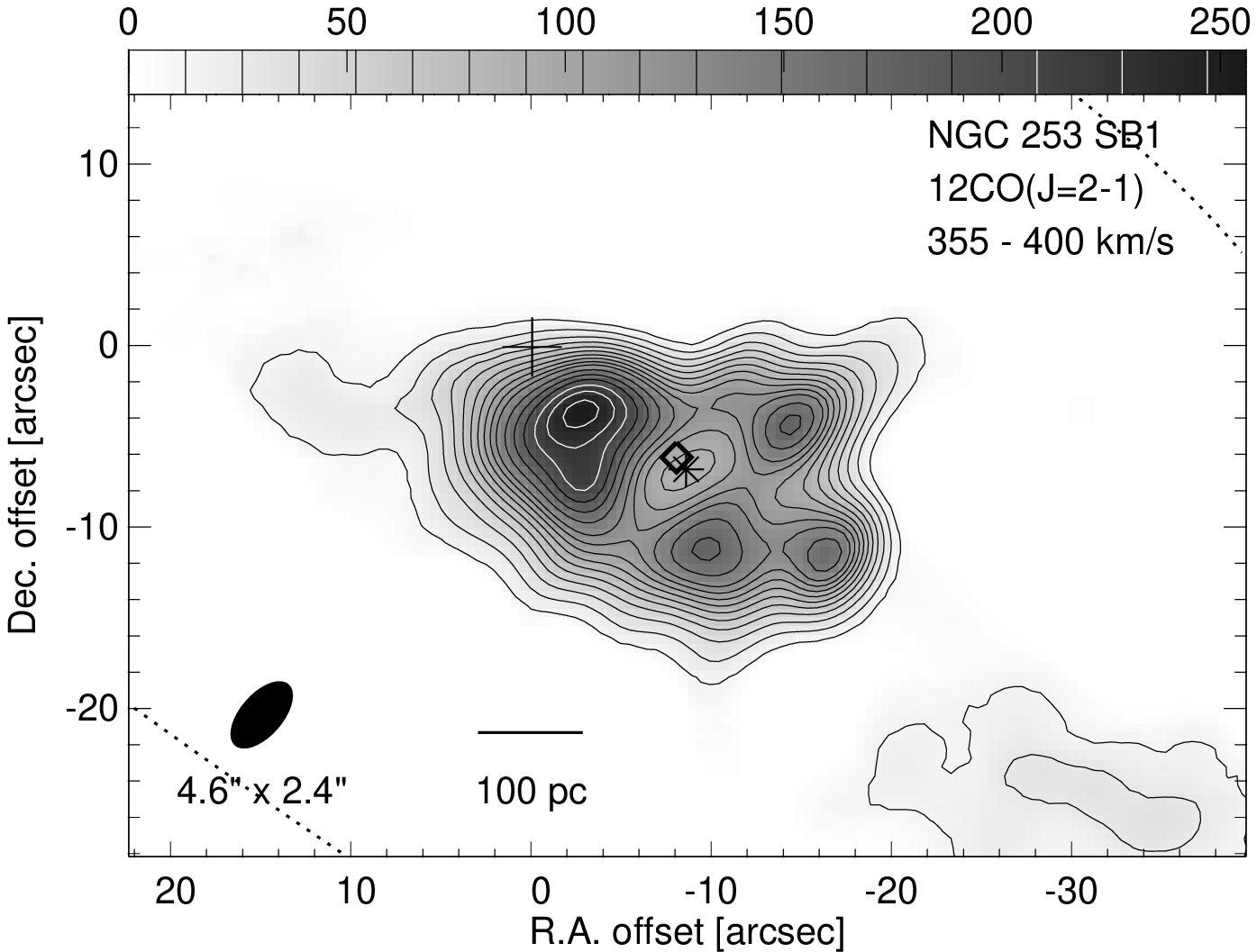}\\
\plotone{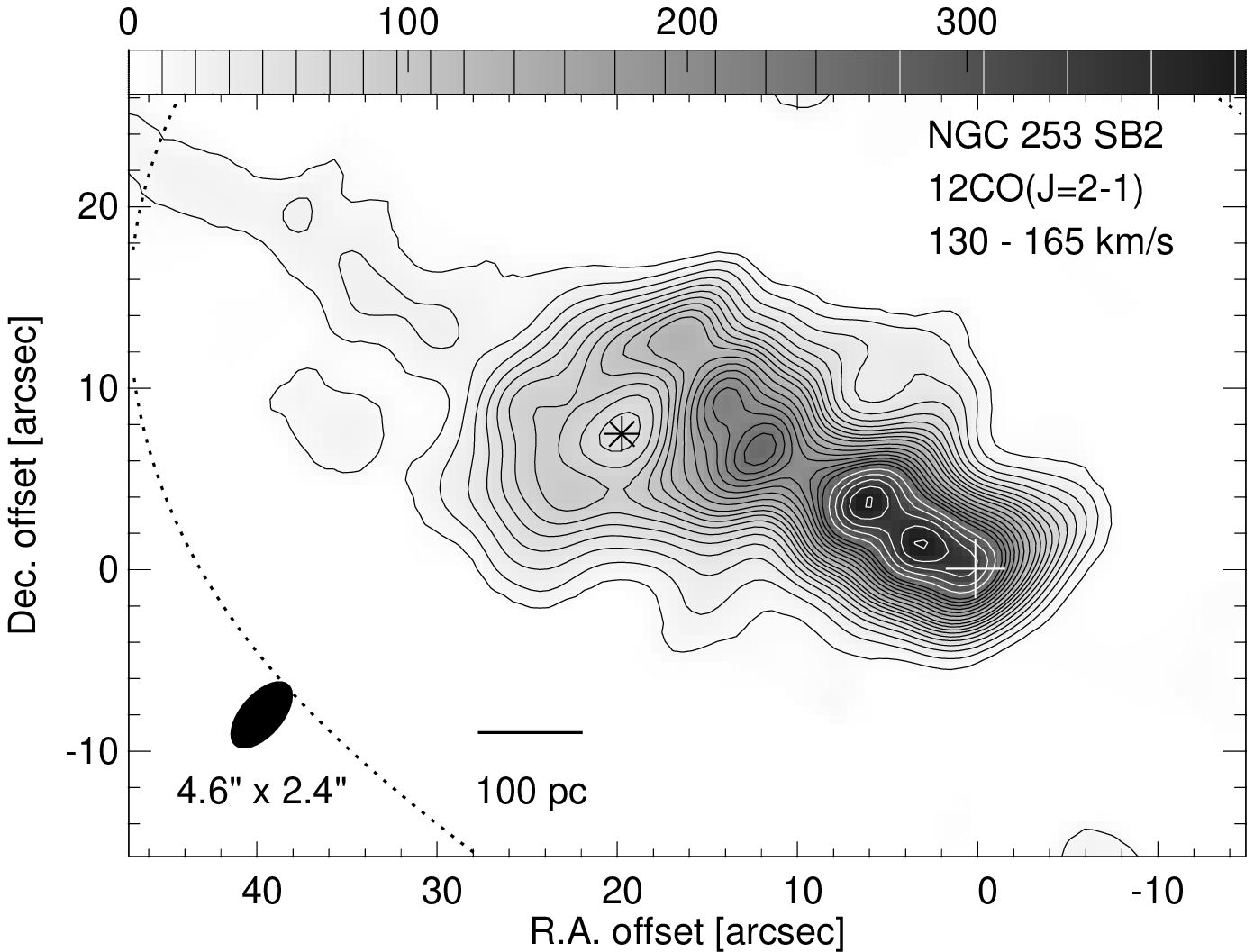} 
\caption{The molecular superbubbles in NGC 253 in \twelveCO(2--1).
The centroid of each superbubble is shown with an asterisk. 
The position of the galactic center, which is the reference position of the maps,  is shown as a cross.
The range of channel velocities used to make each 0-th moment map is shown at the top right corner.
The unit of intensity is Jy \beam\ \kms .
The synthesized beam  is shown at the bottom left corner, 
and the mosaiced primary beam is indicated by a dotted line at the 40\% of its peak.
The diamond in the SB1 map shows the position of the compact radio source called 5.17$-$45.4
in \citet{Ulvestad97}.
 \label{fig.bubbles-integ} }
\end{figure}

The centroid of each shell was determined from the local minimum in the map
that was made by integrating over the central velocities of each shell (Fig. \ref{fig.bubbles-integ}).
Their distances on the sky from the center of the galaxy are 
11\arcsec\ and 22\arcsec\ for SB1 and SB2, respectively.
Their deprojected distances from the galactic center would be 
190 pc for SB1 and 670 pc for SB2 if the centroids are on the midplane of the galaxy. 
SB2 could be at a smaller galactocentric radius if its center is above (i.e., in front of) the midplane of the galaxy; 
a 50 pc vertical offset would place the centroid at the galactocentric radius of 500 pc.  
This may be a more plausible configuration because otherwise the circumnuclear disk would be 
lopsidedly elongated in the direction of SB2. 
Note, however, that the geometrical center of an ISM bubble does not necessarily
coincide with the location of its driving source, 
because a bubble expands faster in the direction of lower ambient pressure.
For SB1, any vertical offset would make its galactocentric radius larger than 190 pc, 
because the center of the feature is almost on the line of nodes.

\subsection{Kinematics, Ages, and Orbits \label{ss-kinematics}}
\begin{figure}
\begin{center}
\epsscale{1.0}
\plottwo{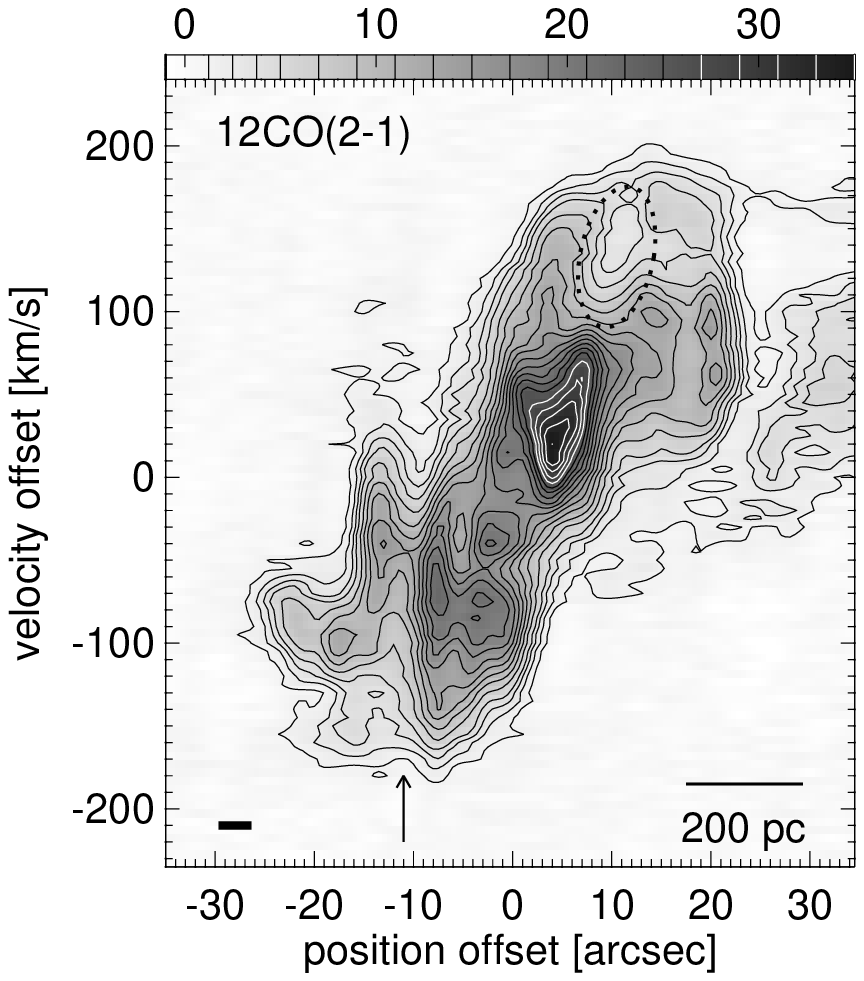}{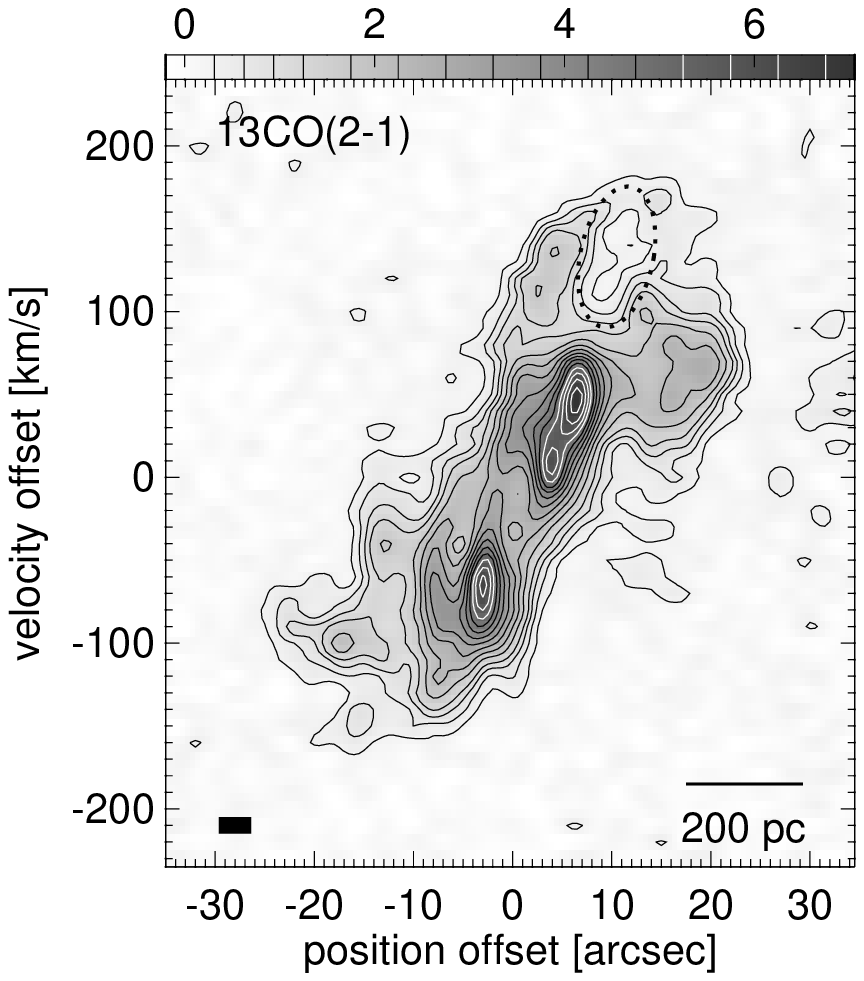}
\end{center}
\caption{Position-velocity diagrams across the superbubble SB1 in the direction parallel to the line of nodes
(P.A.$_{\rm disk} =   231\degr$),
(left) \twelveCO(2--1),
(right) \thirteenCO(2--1).
The intensity is the excess brightness temperature in Kelvin, 
and is corrected for the primary beam attenuation.
The rms noise in the data cube is 0.26 K for \twelveCO\ and 0.10 K for \thirteenCO.
Position offset is measured from the minor axis, 
while velocity is measured from the systemic velocity of 240 \kms.
The dotted ellipse shows the approximate location of the expanding bubble SB1 in each diagram.
The black rectangle at the bottom left corner shows the spacial and velocity resolutions.
The lowest contour is at 4 $\sigma$ for \twelveCO\ and 3 $\sigma$ for \thirteenCO. 
The arrow in the left panel points the trough feature.
 \label{fig.SB1-PV} }
\end{figure}

\begin{figure}
\begin{center}
\epsscale{1.0}
\plottwo{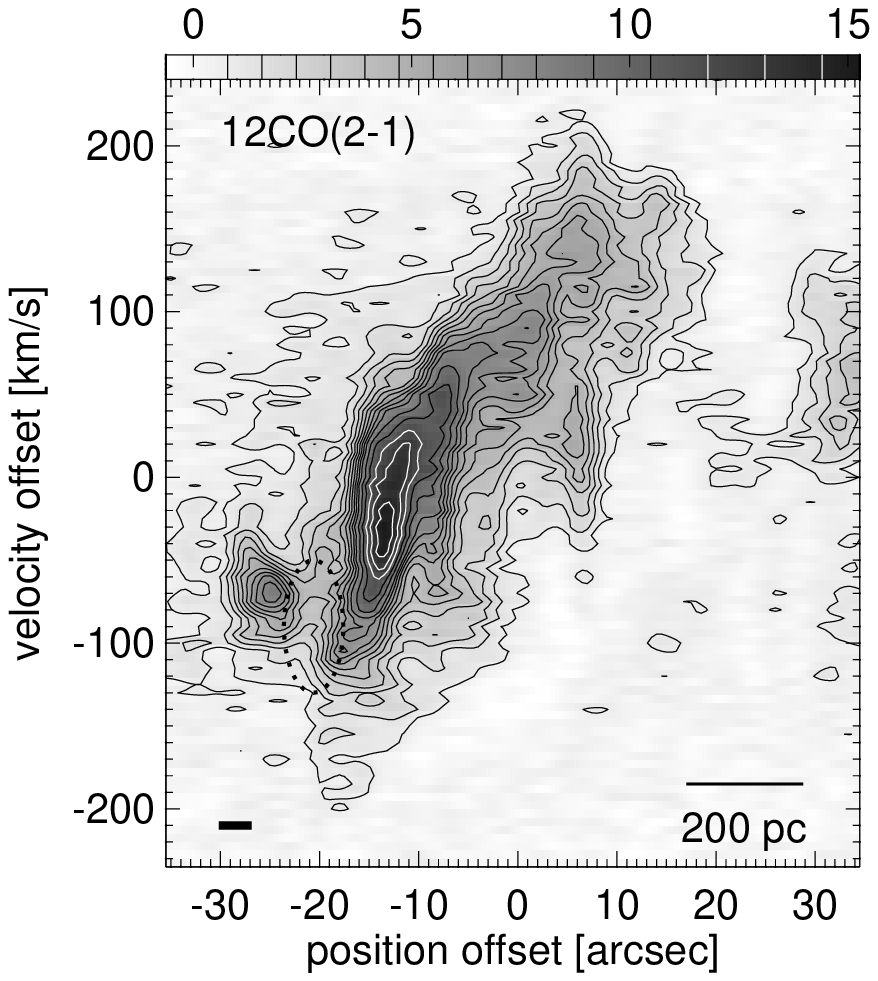}{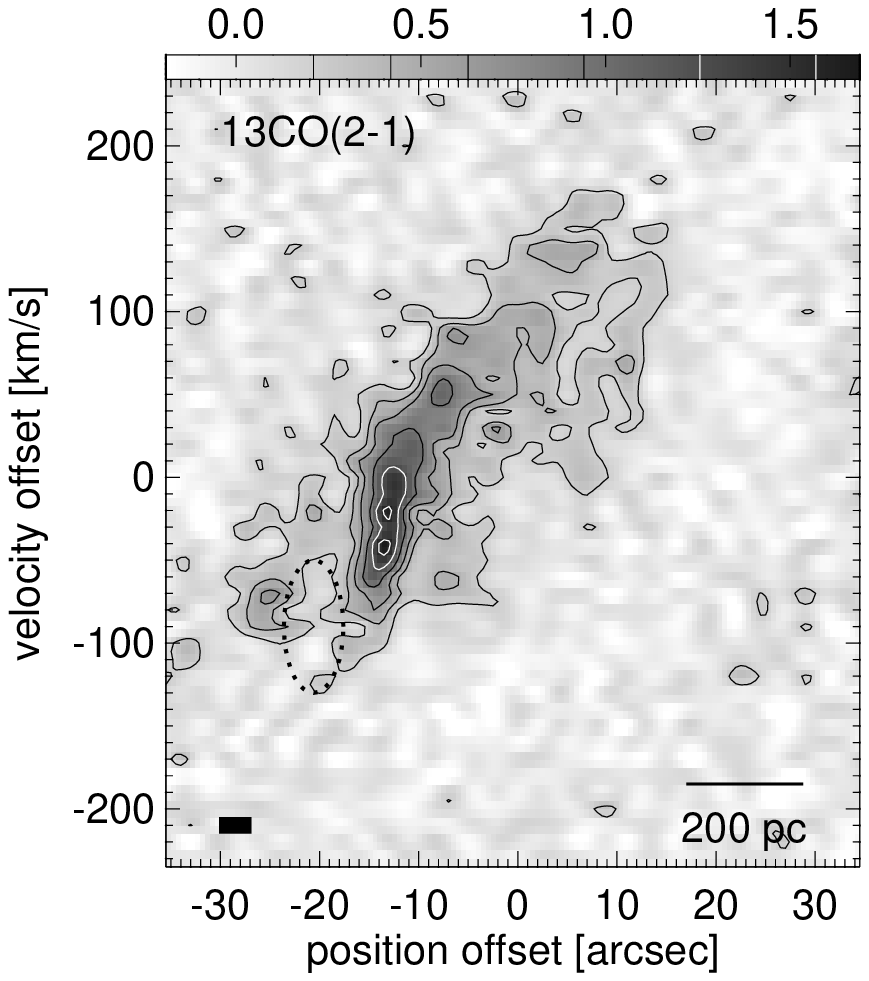}
\end{center}
\caption{Position-velocity diagrams across the superbubble SB2 in the direction parallel to the line of nodes
(P.A.$_{\rm disk} =   231\degr$),
(left) \twelveCO(2--1),
(right)  \thirteenCO(2--1).
The intensity is the excess brightness temperature in Kelvin. 
Position offset is measured from the minor axis, while velocity is measured from the systemic velocity of 240 \kms.
The data are corrected for the primary beam attenuation.
The dotted ellipse shows the approximate boundary of the expanding bubble.
 \label{fig.SB2-PV} }
\end{figure}

The full-velocity widths in which the shells are visible in the channel maps
are about 100 and 80 \kms\ for SB1 and SB2, respectively. 
The velocity structure of SB1 is clearly seen as an elliptical hole or a depression in the
\twelveCO\ and \thirteenCO\ position-velocity (p-v) diagrams in Fig. \ref{fig.SB1-PV}.
The hole structure is similar to the one seen in the p-v diagrams of M82 at the location of the
molecular superbubble in the galaxy \citep[see Fig. 2 of ][]{Weiss99}. 
The shell feature SB2 is less evident  in p-v diagrams; 
a partial hole is marginally visible in \thirteenCO\ 
though it does not  show up as a hole in the \twelveCO\ p--v diagram
 (Fig. \ref{fig.SB2-PV}).
A hole in a p-v diagram suggests that the shell is around a three-dimensional cavity 
that is either expanding or contracting.
The latter is highly unlikely because a coherently contracting shell of \tilde100 pc size is
difficult to create, whereas an expanding shell can be produced around a point explosion.
The expansion caused by an explosion would be faster along the minor axis of the galaxy 
because of the decreasing pressure away from the disk. 
The elongation of the bubbles in Figs. \ref{fig.bubbles-channel} and \ref{fig.bubbles-integ}, 
if confirmed, would be consistent with this expectation.

Assuming that the shells are expanding bubbles, the kinematical age of each shell is estimated
to be $\approx 0.5$ Myr.
This is derived from $t_{\rm kin} \approx \alpha R/v$, 
where $R$ is the radius and $v$ is the expansion velocity.
We use the semi-minor axis  and half velocity width of each bubble for $R$ and $v$ respectively,
so that both are the ones in the plane of the nearly edge-on galaxy.
Cylindrical symmetry of each feature is assumed.
The parameter $\alpha$ is to account for non-linear expansion in the form of $R\propto t^{\alpha}$. 
For example, $\alpha$ is 1 for free expansion, 
3/5 for adiabatic expansion due to constant energy injection,
2/5 for non-radiative expansion due to a single explosion (the Sedov-Taylor phase), 
2/7 for the pressure-driven expansion of a shell around a hot bubble,
and
1/4 for momentum-conserved snowplow expansion \citep[see, e.g.,][]{Ostriker88};
all the explosions are assumed to take place in a 
uniform interstellar medium with negligible pressure.
The power $\alpha$ could be larger than 3/5 if the energy injection is continuous and has
been increasing with time, as one might expect from a gradually forming star cluster. 
We adopt $\alpha\approx 0.5$ as a value in the middle of possible values
and expect a factor of 2 uncertainty in the estimated age.

We can also infer orbital properties of the bubbles from their mean velocities and the p-v diagrams.
There are basically two types of oval orbits for gas and stars in a galaxy bar. 
Namely, the dominant \xone\ orbits are elongated along the bar, 
and  \xtwo\ orbits elongated perpendicular to the bar exist near the galactic center under certain circumstances
\citep[c.f.,][]{Binney87}.
Gas can take intermediate orbits near the transition region between the two types of orbits.
In the configuration of NGC 253, a gas cloud in an \xtwo\ orbit should have larger line-of-sight velocity 
than that in an \xone\ orbit, as modeled by \citet{Peng96}.
The presence of two types of orbits (i.e., velocity components) in the central \tilde50\arcsec, 
where there is the nuclear gas disk, 
is clearly seen in the single-dish CO p-v diagram of \citet[their Fig. 17(a)]{Sorai00}.
The high-velocity component there is most likely in the \xtwo\ orbits 
or in the orbits that are intermediate between \xone\ and \xtwo, 
whereas the low-velocity component within about 50 \kms\ from the systemic velocity 
must be in the \xone\ orbits.
The bubble SB1 and its progenitor should belong to an \xtwo\ orbit
because the mean velocity of the bubble is close to the terminal velocity in the p-v diagram
and about 130 \kms\ away from the systemic velocity.
The association of SB2 to another \xtwo\ orbit is likely but less certain,
because the offset from the systemic velocity is 90 \kms. 
The ages of the bubbles are several ten times shorter than their orbital periods around the galactic center. 
Thus, differential rotation could not have significantly affected the bubbles.

\subsection{Energetics}
The total energy needed to create each bubble is estimated in two ways
to be \tilde$10^{46} $ J
from the size and expansion velocity of the bubble 
as well as the mean density of the ambient gas.
First, the numerical simulation of a point explosion in the uniform interstellar medium 
by \citet{Chevalier74} gives 
the following formula for the total energy of the single explosion that caused the bubble: 
\begin{equation}
	 \frac{E}{10^{43} {\rm J}}
	= 
	5.3 \times 10^{-7} 
	\left(\frac{n_{0}}{\rm cm^{-3}} \right)^{1.12}
	\left(\frac{v}{\rm \kms} \right)^{1.40}
	\left(\frac{R}{\rm pc} \right)^{3.12},
\end{equation}
where $n_{0}$ is the hydrogen number density of the pre-bubble medium. 
Alternatively, one can use the similarity solution of a bubble caused by continuous energy injection
into the interstellar medium.  
The total injected energy,  
20\% of which goes to the kinetic energy in the solution \citep{Weaver77}, 
is estimated to be
\begin{equation}
	E \approx \frac{10 \pi}{3} \rho_{0} v^2 R^3,
\end{equation}
where $\rho_{0}=1.4 n_{0} m_{\rm H}$ is the initial mass density  of the ISM.
The gas density in both formulae is the one prior to the bubble formation and is assumed  to be uniform.
One therefore has to estimate the pre-bubble gas density averaged
over the volume that  the bubble currently occupies.  
We adopt  $n_{0} \sim 70 $ \percubiccm\  and  $\sim 20 $ \percubiccm\ for SB1 and SB2, respectively, 
on the basis of the gas column densities estimated for the nuclear gas disk
\citep{Wall91,Mauersberger96,Harrison99,Sorai00}. 
Specifically, we used  
\twelveCO(2--1) integrated intensities of $2\times10^3$ and $0.5\times10^3$ K \kms\ for the bubbles, 
the inclination of the disk of 78\degr,
$N_{\rm H_{2}} / I_{\rm CO}=0.3\times10^{20}$ \persquarecm\  (K \kms )$^{-1}$, 
and 
a 100 pc height for the bubbles. 
The kinetic energy needed to create the $R\sim 50$ pc bubbles is estimated to be
 2--3$\times 10^{46}$ J for SB1 and
0.4--0.6$\times 10^{46}$ J for SB2.
The smaller numbers are from the similarity solution, in which radiative loss of energy is ignored.
The mass of each shell is $M_{\rm shell} \sim 1.3\times10^{6}$ \Msol\ and $0.4\times10^{6}$ \Msol\ for SB1 and SB2,
respectively.

It is noted that the error of the derived energy may be as large as an  order of magnitude 
in the worst case.
This is because the energy depends almost linearly on the ambient gas density, 
whose uncertainty is no less than a factor of a few,  
and also depends on many parameters that are not in the models, such as non-uniformity
of the ambient gas.
The pressure of the ambient gas is among those ignored in the models even though the
pressure 300 pc from the center of NGC 253 has been estimated to be \tilde$10^{-10}$ Pa
\citep[i.e., $P/k \sim 10^7$ \cubiccm\ K; ][]{Heckman90,Carral94}, 
which is about 2 orders
of magnitude higher than the pressure in the center of our Galaxy \citep{Sawada01}.
The continuous energy-injection model gives the internal pressure of a bubble as
$P_{\rm in}=\frac{7}{3}\rho_{0} v^{2}$, and each bubble has  \tilde5 times higher 
internal pressure than external pressure according to the formula.
Thus the bubbles probably have not been affected significantly by the external pressure yet.
However, now that there are morphological hints of blowouts and  the size of the bubbles 
is indeed comparable to the thickness of the molecular gas layer in galactic centers 
\citep[][and references therein]{Scoville93b}, 
the bubbles are likely to be approaching the end of the pressure-driven expansion 
in the disk plane of the galaxy.

The energy needed to create each bubble, \tilde$10^{46}$ J, 
corresponds to the kinetic energy released by \tilde100 supernovae (SNe),
or \tilde1--10 hypernovae (HNe) each of which releases 10--100 times more kinetic energy  than a
supernova \citep{Iwamoto98, Nomoto04}.
This suggests that, unless the energy came mostly from a hypernova,  
each of the expanding bubbles was 
created by successive or continuous injection of energy rather than by a single explosion.
The formation of a large ISM bubble by continuous energy injection is consistent with 
the model of superbubbles, which are created by stellar winds
and supernova explosions of massive stars in a star cluster \citep{Tomisaka81, McCray87}.
It is also reasonable to call an ISM bubble caused mainly 
from a hypernova explosion a superbubble, 
because the involved energy matches that of multiple supernovae 
in the original superbubble model.
Thus, SB1 and SB2 can be called superbubbles in terms of energetics.

\subsection{Association with Features in other Wavelengths}
There is only one compact radio source inside the bubbles 
according to the list of centimeter-wave continuum sources
of \citet{Ulvestad97}, even though about 60 compact sources most likely 
comprised of supernova remnants
and HII regions have been detected in the nuclear region.
Most of the radio sources are concentrated around the starburst nucleus between the two bubbles.
The  radio source listed as 5.17$-$45.4 
is within 1\arcsec\ from the center of SB1 (see Fig. \ref{fig.bubbles-integ})
and has a 6 cm flux density of 0.32 mJy \citep{Antonucci88}.  
The source by no means stands out among the radio sources in the nuclear region, nor is
it very prominent in terms of its absolute radio power.
The radio power of the source, $P_{\rm 6 cm}=5\times10^{17}$ W Hz$^{-1}$, 
is equivalent to that of Cas A, 
a single supernova remnant in the Galaxy.
The lack of a significant population of centimeter sources in the bubbles, however, 
does not mean that there have been few supernovae.
This is because the VLA observations can detect a supernova remnant
for only several 100 years, as can be estimated from the supernova rate, 0.1 \peryr, and
the number of radio sources detected in the starburst region.
For a comparison, the mean interval between supernovae in the bubble would be 5000 yr 
to have 100 SNe in 0.5 Myr.
The bubbles do not coincide with known OH and H$_2$O masers, 
which are concentrated to the central 200 pc of the starburst nucleus 
\citep{Frayer98, Henkel04}.

There is a faint spot near the center of SB1 
in the HST image of the [FeII] 1.644 \micron\ emission line \citep[Fig. 5b]{Alonso-Herrero03}. 
The spot is one of the compact sources that show spatial correspondence with a centimeter
radio source.
Compact [FeII] sources in a starburst are thought 
to trace supernova remnants for \tilde$10^4$ yr.
In the HST study, the median [FeII] luminosity of the compact sources  
is two orders of magnitude larger than that of galactic supernova remnants.
Thus the sources may well be clusters having multiple supernova remnants,
though higher density of the ISM could also boost the line luminosity \citep{Morel02}.
In X-rays, inspection of Chandra images reported in \citet{Strickland00} and \citet{Weaver02}
did not reveal any obvious source in the bubbles.

The two superbubbles are around the waist of the superwind,
which is a kiloparsec-scale conical outflow of ionized gas from the starburst region 
into the direction perpendicular to the galactic disk.
The base of the superwind has a diameter of \tilde20\arcsec\
in X-ray, near-IR \HH\ line, and synchrotron emission \citep{Strickland00, Sugai03, Ulvestad97}.
The bubble SB1, which is 11\arcsec\ from the galactic center, is almost at the surface of the biconical outflow.
The CO tongue extending northwest from around SB1 corresponds to the \HH\ spur B of \citet{Sugai03} 
and also coincides with the foot of the most prominent spur in the 20 cm image of \citet{Ulvestad97}.
There are two possible causes for the CO tongue if it is a \tilde200 pc
out-of-plane structure as we inferred.
One possibility is a blowout from the superbubble SB1 accidentally overlapping 
with the superwind cone,
and the other is molecular gas entrained by the large-scale superwind.

\subsection{Gas Properties in and around the Bubbles}
\begin{figure}
\epsscale{0.5}
\plotone{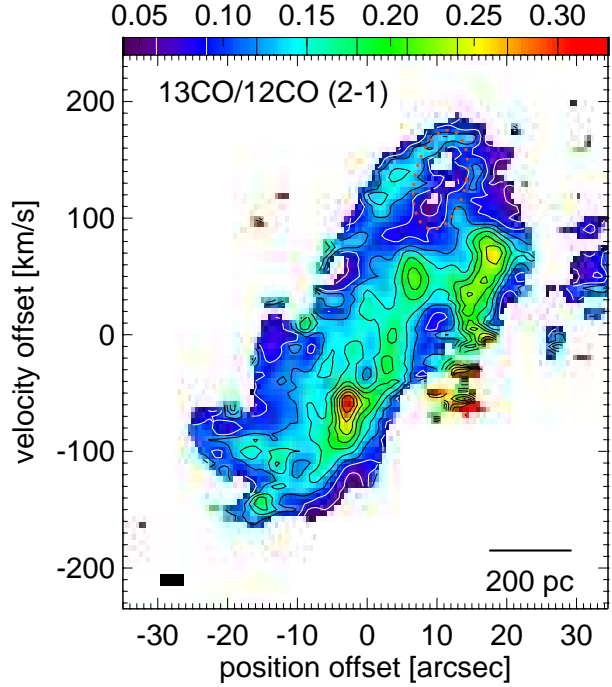}
\caption{Position-velocity diagram of  the \thirteenCO(2--1)-to-\twelveCO(2--1)  ratio
of excess brightness temperature.
The p-v diagram is across the superbubble SB1 in the direction parallel to the line of nodes
(P.A.$_{\rm disk} =   231\degr$). 
The ratio was calculated by using maps corrected for the primary beam attenuation
and convolved to the same resolution.  
The ratio was calculated only where both lines are detected above 2.5 $\sigma$. 
The black rectangle at the bottom left shows the spacial and velocity resolutions.
The ratio approximately corresponds to the optical depth of the \thirteenCO(2--1) line.
Note the low opacity gas in and around the superbubble, which is
marked by an ellipse in a dotted line.
 \label{fig.SB1_color-PV} }
\end{figure}

A superbubble is expected to affect the state of the ISM in and around the expanding shell.
In an attempt to see the effect, we made a position-velocity diagram  through SB1 for the
\thirteenCO-to-\twelveCO\ temperature ratio.
The p-v diagram shown in Fig. \ref{fig.SB1_color-PV} indicates that the line ratio
in the bubble region is lower by a factor of a few than the ratio at the ambient non-bubble locations.

There are a number of possible reasons for the lower ratio at the bubble location.
The lower ratio can be simply due to the lower column density per unit velocity in the bubble, 
because that reduces \twelveCO\ opacity.
Also, the line ratio can be due to the excitation conditions of the CO lines.
In the bubble, the number density of molecular gas may be 
below the critical density to excite CO to the J=2 level. 
It would make the excitation of \thirteenCO\
much harder than that of \twelveCO\ 
because photon trapping does not help the optically thin \thirteenCO.
Alternatively, the gas temperature in the bubble can be high enough to move most population
above J=2, making both \twelveCO\ and \thirteenCO\ optically thin, and reducing the intensity 
ratio to the abundance ratio of the two molecules.
In addition, selective photodissociation of \thirteenCO\ molecules \citep{vanDishoeck88}
may result from intense UV radiation in the bubble, again reducing the line ratio.
Observations of other transitions and molecules would help to distinguish between these possibilities.

\section{Evolution of the Superbubbles, and Their Progenitors}
We have seen that the \tilde100 pc shell features found in the molecular gas around
the nuclear starburst of NGC 253 can be modeled as expanding superbubbles 
of ages \tilde0.5 Myr and injected kinetic energies of \tilde$10^{46}$ J.  
The evolution and progenitors of such superbubbles can be inferred from the current data.

\subsection{Young Bubbles and their Evolution} 
The kinematical ages of the superbubbles, \tilde0.5 Myr, are rather short compared to the
age of the starburst,  which is estimated to be  20--30 Myr \citep{Engelbracht98}.
Why are older superbubbles missing?

One of the possible solutions for this mystery is 
that older bubbles are difficult to find because they must have blown out of the
galactic plane and have ceased expansion.
If a superbubble similar to the ones we found has reached a point of blowout, 
as at least SB1 seems to have, 
then its expansion within the galactic plane loses the thrust from internal pressure
and becomes a momentum-driven snowplow.
The expansion in the plane of the galaxy halts 
(i.e., the expansion velocity reduces by a factor of 5 and becomes comparable
to the turbulent velocity of the ambient molecular gas) in about 2 Myr.
The size of the bubble increases only by about a factor of 2 in that time. 
After that the bubble will no longer have a quasi-spherical shape nor significant expansion velocity
when viewed from edge-on.
Thus such a superbubble would be difficult to recognize despite its large size.
An older bubble would get less recognizable 
also because its outer wall becomes fragmented 
and because its shape in the galactic plane is deformed from the original circular one
owing to the differential rotation and non-circular gas motions in the region.  
If a typical molecular superbubble could be detected for only a few Myr with one's resolution and sensitivity,  
then it may well happen that one finds two half-Myr old superbubbles without seeing older ones.

It is also possible that star formation in the radii where we found the superbubbles 
has been retarded 
compared to the nuclear starburst inside, reaching the stage of superbubble formation only recently.
The delay could be due either to the inside-out progression of star formation or to the
time lag statistically expected from the low superbubble formation probability in the outer region.
In contrast, the nuclear region with a \tilde$10^{7}$-yr-old starburst may be 
already too disturbed or have too high ambient pressure to form a new distinct
superbubble. 
The energies from stellar winds and supernovae there may be channeled almost 
directly to the superwind.

We note that the molecular superbubble found in M82 is also very young, 
having a $\sim$1 Myr age \citep{Weiss99,Matsushita00}.
Being young (only about 1 Myr old) is a common character of the molecular 
superbubbles found so far in two of the nearest starburst galaxies. 
This may well be due to the short time window of detection for superbubbles 
in edge-on galaxies,
because both starburst galaxies are seen nearly edge-on. 

\subsection{Super Star Clusters}
The large energy and the young age of each superbubble provide a strong constraint
on the progenitor, because a large mechanical luminosity is required to create the bubble.
The progenitor of each bubble must have had a mean mechanical luminosity 
of $1\times10^{33}$ W in the last 0.5 Myr in
a region smaller than 100 pc.
This luminosity can be provided by a $10^{6}$ \Msol\ star cluster that has the Salpeter initial mass function,
an upper mass cutoff of 100 \Msol, and solar metallicity,
according to the Starburst99 model of \citet{Leitherer99}.
In the case of instantaneous star formation in the cluster, 
the luminosity stays roughly constant for about 30 Myr,
first coming from stellar winds and then from supernovae.
The duration of successive supernova explosions,  \tilde30 Myr, is set by the
range of the lifetimes of the massive stars (\tilde8--100\Msol) that eventually explode as core-collapse supernovae.
The model cluster at the age of 1 Myr has 5000 O stars, 
absolute K-band magnitude of $M_{K}\approx-16$ mag,
and bolometric luminosity of $1\times10^{9}$\Lsol, or  3 \% of the total far-IR luminosity of 
the starburst and that of the galaxy
\citep[$L_{\rm FIR}=(3.0$--$3.5) \times10^{10}$\Lsol,][]{Telesco80,Sanders03}.
A star cluster of this magnitude, if it exists, must be a significant building block 
of the starburst in NGC 253.

The mass of the cluster postulated here agrees with that of the super star clusters 
found in many starburst galaxies including NGC 253 
\citep[and references therein]{Whitmore03},
though the cluster mass, as well as the numbers derived from it,
has uncertainty inherited from the uncertainties in the expansion energy
and age of the superbubbles.
Super star clusters are therefore possible driving sources of the superbubbles.
The measurements of the size and stellar population of the progenitor clusters, if any,
are needed to conclude the identification.

At the distance of NGC 253, the cluster's magnitude will be $m_{K}\approx12$ mag without extinction,
according to the Starburst99 model. 
Extinction toward the center of each bubble is estimated from average \HH\ column densities 
to be $A_{K}\lesssim $ 5 and $\lesssim 2$ mag for SB1 and SB2 respectively, 
though small scale structure in the molecular disk may considerably change local extinction in our lines of sight. 
Thus the clusters would nominally be at 14--17 mag in the K band.
The bubble locations have $K_{s}$-band surface brightness of 14--15 mag arcsec$^{-2}$
in the 2MASS Large Galaxy Atlas \citep{Jarrett03}. 
The 2\arcsec\--3\arcsec\ resolution of the data, however, is insufficient to identify compact sources in the region.
The high-resolution near-IR images of  \citet{Alonso-Herrero03} do not show a conspicuous source 
at the bubble locations in continuum, though a source is seen in SB1 in the  [Fe II] line as we already noted.
The datasets we found in the literature  do not allow us to set a constraint in the K band.
In 1 mm continuum, the uncertainty in dust configuration in our lines of sight to the progenitors 
makes it difficult to set a constraint.
On one hand, a cluster of $1\times10^{9}$\Lsol\ should have been detected 
at \tilde30 mJy, which is \tilde7$\sigma$ in our continuum map, if dust reprocesses the cluster luminosity 
by the same degree as it does to the bulk of the starburst luminosity in NGC 253.
On the other hand, the conversion to dust emission may be less effective for the progenitors 
considering that they should be in cavities of interstellar medium.

If the two superbubbles are indeed made by star clusters, then the bubbles must have
spent only a few \% of their energy injection period until now.
In other words, the total kinetic energy eventually released inside each superbubble through SNe would be 
 \tilde$10^{48}$ J, or several 10 times larger than the energy deposited so far.
Because a superbubble blows out of the disk early, 
most of the energy will be channeled to the galactic halo as depicted
in the galactic chimney model by \citet{Norman89}.
In a similar starburst in M82, several chimneys are indeed seen in deep radio and 
optical images \citep{Wills99}.
The detection of such chimneys in NGC 253, however, would be more difficult  
because NGC 253 is not completely edge-on.

\subsection{Hypernovae}
A hypernova can, as we saw, provide most of the mechanical energy needed to form each of the superbubbles.
It also satisfies the constraints on the mean mechanical luminosity and compactness
because it is an instantaneous point explosion.
Furthermore, it is possible from a statistical viewpoint that the observed molecular superbubbles 
are due to a couple of hypernovae.
The supernova rate in the nuclear starburst in NGC 253 is estimated to be about 0.1 \peryr\ 
\citep{Rieke88,vanBuren94,Ulvestad97}.
The hypernova-to-supernova ratio, on the other hand, has been estimated to be $10^{-3}$--$10^{-5}$ \citep{Paczynski98,Podsiadlowski04}.
Thus the hypernova rate in the starburst region of NGC 253 is
expected to be in the range of \tilde$10^{-5\pm1}$ \peryr.
This is comparable to or larger than the superbubble formation rate of $10^{-6}$ \peryr\ estimated from the
current number of superbubbles, namely 2, and the time scale they are recognizable, \tilde2 Myr.
It is therefore possible, not only energetically but also from the formation rate argument, that the
superbubbles were caused by hypernova explosions. 
Note that there must be 1--100 hypernova remnants of less than 1 Myr old in the starburst region,
regardless of whether the two superbubbles were due to hypernovae.

The total kinetic energy released from the hypernova population is much less than that from 
supernovae, when integrated over the  starburst region and over the entire duration of the starburst. 
It is because the larger number of supernovae more than compensates for the smaller energy
of each explosion.
However, supernovae can create a superbubble only if they are concentrated both in space and time,
as in a star cluster.
This constraint does not apply to hypernovae 
so long as the superbubbles that could be created with a single hypernova explosion are concerned.
Therefore hypernovae can be more significant than expected from
the total-energy argument as a mechanism 
to create molecular superbubbles in an  environment without massive clusters.
On the other hand, a hypernova in an environment with many massive clusters
is more likely to occur in one of those clusters and to cooperate with stellar winds 
and supernovae to form a superbubble around the cluster.

Observationally, the largest difference between a super star cluster and a hypernova as the
progenitor of a molecular superbubble  is 
that the former must be still present as a cluster of thousands of OB stars.
If such a cluster is not found in a superbubble in any wavelengths,  
then the progenitor is more likely to be a hypernova.

Regarding the effect on the ISM,  a hypernova will have 
less lasting impact than a super star cluster, even though each forms a similar
superbubble when compared at the age of \tilde1 Myr.
The difference is because, as we noted, massive stars in a super star cluster last for a few 10 Myr
and release \tilde100 times more kinetic energy than a hypernova.

\section{Discussion}
Our observations revealed many gas properties and features that are most likely caused
by the starburst. 
They include the high gas temperature, 
superbubbles, and molecular gas likely entrained by the superwind.

Regarding the bubbles, we saw that a number of properties of the two shell features 
are consistent with them being  expanding superbubbles.
To be precise, however, that does not {\it prove} them to be superbubbles.
Cold gas in a barred galaxy has non-circular motion and non-axisymmetric distribution
due to the bar, 
and it could show a bubble-like feature in a p-v diagram, as pointed out by \citet{Wills00}.
We therefore tested if we see the shell-like features in gas dynamical models in various barred potentials, 
by taking the models of \citet{Athanassoula92} and plotting them in the same way as NGC 253 was observed.
The bubble features in the sky-plane maps and p-v diagrams could not be reproduced in the models we tested. 
In particular, the bar models have an inherent difficulty to explain two shell-like features 
whose positions are not symmetric about the galactic center, 
because bar-induced features tend to be bisymmetric. 
Some additional factors, such as  $m\neq 2$ modes in the bar potential and self-gravity of gas, 
are required to break the bisymmetry.
We have not explored such models, because the parameter space becomes vastly larger for them. 
In this regard, there remains an untested possibility that factors unexplored by us conspired in NGC 253
to create gas features masquerading as expanding bubbles.
We just note that such gas features, if confirmed, would be rarer findings than the molecular superbubbles
that have been reported in several galaxies 
\citep[see e.g.,][and those already cited for the molecular superbubble in M82]{Handa92, Irwin96, Rand00, Walter04}.
We conclude that the current data most favor the superbubble model.
Further observations, in particular the ones to look for the progenitors or their remnants in various wavelengths, 
are desirable.

Unusual features in the central molecular disk in NGC 253 are not limited to the couple of  gas bubbles. 
The \twelveCO\ position-velocity diagram shown in Fig. \ref{fig.SB1-PV} is highly disturbed
in comparison with those in the collection of CO p-v diagrams in \citet{Sakamoto99}. 
For example, the trough at the offset of $-11$\arcsec\ and the emission components 
on both of its sides having nearly 150 \kms\ velocity widths are quite unusual.
The gas dynamical simulations we mentioned above did not produce the trough feature either. 
These features may be related to the superwind cone around that location or may be due to another superbubble, 
a very young one that would be characterized by its compact size and large velocity width.

The numerous signs of disturbances in the central disk once
again prove NGC 253 to be one of the best targets to study starburst.
The features we found are of smaller scales (i.e., 100 pc or less) than the ones previously known, thanks to
the higher resolution of our observations.
These kinematical and thermal inputs to the local interstellar medium should almost certainly influence 
the subsequent star formation in the region, through the formation and destruction of star forming gas clouds. 
These inputs should therefore affect the evolution of the starburst. 
Details of this  process remain to be seen in further studies. 

\section{Summary}
We observed molecular gas and dust in the central \tilde2 kpc of the archetypal 
starburst galaxy NGC 253 at resolutions of $\sim$60 pc and 5--10 \kms. 
Our main conclusions are the following:
 
\begin{enumerate}
\item Molecular gas in the central region is mostly distributed in the nuclear gas disk
of \tilde500 pc radius. 
The gas in the disk is warm, with a peak brightness temperature
of \tilde40 K in \twelveCO\twotoone.  The area-filling factor of the gas appears high in the center of the disk, suggesting the observed brightness temperature to be close to the gas temperature. 
There are hints that low density molecular gas is flaring above the midplane possibly due
to the superwind, and that the ISM in the central 100 pc has distinctive physical conditions.

\item We have found two molecular shells or cavities of \tilde100 pc size in the nuclear gas disk. 
These features show velocity structure indicative of expansion at a velocity of \tilde50 \kms.

\item We modeled these features as molecular superbubbles. 
In the model, each feature has  an age of \tilde 0.5 Myr, involves \tilde$10^{6}$ \Msol\ of interstellar medium,
and needed the kinetic energy of \tilde$1\times10^{46}$ J for its formation.

\item The large amount of energy deposited within the short timescale calls for
a progenitor with a high rate of energy injection, $1\times10^{33}$ W averaged over
the age of the superbubbles.
A \tilde$10^{6}$ \Msol\  star cluster of \tilde1 Myr age would have
enough mechanical luminosity to create
such a superbubble with the stellar wind and from supernova explosions of 
thousands of OB stars. 
If such a cluster had a small size of $\leq 10$ pc, it would be a super star cluster.
It could contribute a few percent of the total bolometric luminosity of the starburst in NGC 253, 
and would be a significant building block of the starburst.

\item Alternatively, a hypernova, which releases 10--100 times more
kinetic energy than a supernova,  
could be a significant source of energy for each superbubble.
The hypernova rate in the starburst region, inferred from the supernova rate there, 
is of the same order of or larger than the formation rate of the superbubbles.
Thus a hypernova is a tenable progenitor of each superbubble.

\item Our data suggest that the nuclear starburst in NGC 253 had strong thermal and kinematical impacts 
on the central molecular disk hosting the starburst. 
In particular, our high-resolution observations have revealed various signs of disturbance 
at 100 pc or smaller scales in the gas disk.  
These may well influence the formation and destruction of star forming gas clouds, 
and thereby affect the evolution of the starburst. 

\end{enumerate}

\acknowledgments
We thank the SMA team members whose hard work made the
observations reported here possible.
We also thank Dr. Tsuyoshi Sawada for providing us with information on his CO(2--1) data of the
Galactic center,
Dr. Alumudena Alonso-Herrero for providing us with her HST images of NGC 253,
and Dr. Keiichi Wada for providing us with his numerical simulation codes.
This research has made use of NASA's Astrophysics Data System Bibliographic Services,
and has also made use of the NASA/IPAC Extragalactic Database (NED) 
which is operated by the Jet Propulsion Laboratory, California Institute of Technology, 
under contract with the National Aeronautics and Space Administration.


\newpage

\begin{deluxetable}{ccc}
\tablewidth{0pt}
\tablecaption{Molecular Superbubbles in NGC 253 \label{table.bubblesl}}
\tablehead{\colhead{parameter} & \colhead{SB1} & \colhead{SB2} \\
\colhead{ } & {(J0047325$-$251724)} & {(J0047346$-$251709)}  }
\startdata
$\alpha_{\rm J2000}$ &             00\hr47\mn32\fs53               &              00\hr47\mn34\fs64       \\
$\delta_{\rm J2000}$  &  \minus25\arcdeg17\arcmin24\farcs0       &  \minus25\arcdeg17\arcmin09\farcs5        \\
$V_{\rm LSR}$ [\kms]  &             375    					     & 150 \\
size   [arcsec]  \tablenotemark{a}   &   $9 \times 6 $       &     $10 \times 6 $     \\  
major axis P.A.  \tablenotemark{a} & 135\degr  & 135\degr \\
$\Delta V$ [\kms]  &   100       &      80   \\
$t_{\rm kin}$ [Myr] \tablenotemark{b} &  0.5 &  0.6  \\
$R_{\rm G.C.}  $ [pc]  \tablenotemark{c} & 190 & $\lesssim$ 500 (or 670)\\
$E$ [$10^{46}$ J]  & 2  & 0.4 \\
$M_{\rm shell}$ [$10^{6}$\Msol] & 1.3  & 0.4 
\enddata
\tablenotetext{a}{approximate size and position angle. 
See text for details. 1\arcsec\ = 17 pc.}
\tablenotetext{b}{Kinematic age.}
\tablenotetext{c}{Distance from the galactic center. 
See text for the discussion on the two possible assumptions about SB2.}
\end{deluxetable}

\clearpage

\clearpage

\end{document}